%% file: main.tex
\documentclass[letterpaper,twocolumn,10pt]{article}
\usepackage{usenix2019_v3}

\usepackage{tikz}
\newcommand*\ec[1][1ex]{\tikz\draw (0,0) circle (#1);} 
\newcommand*\hc[1][1ex]{%
  \begin{tikzpicture}
  \draw[fill] (0,0)-- (90:#1) arc (90:270:#1) -- cycle ;
  \draw (0,0) circle (#1);
  \end{tikzpicture}}
\newcommand*\fc[1][1ex]{\tikz\fill (0,0) circle (#1);} 

\newcommand{\tk}{\ding{52}\xspace}
\newcommand{\cx}{\ding{56}\xspace}

\usepackage{bigstrut}
\usepackage{enumitem}
\usepackage{amsmath}
\usepackage{graphicx}
\usepackage{pict2e}
\usepackage{callouts}
\usepackage{caption}
\usepackage{subcaption}
\usepackage{listings}
\usepackage{booktabs}
\usepackage{multirow}
\usepackage{xcolor}
\usepackage{colortbl}
\usepackage{xurl}
\usepackage{xstring}
\usepackage{arydshln}
\usepackage{circledsteps}

\usepackage{xspace}
\usepackage{numprint}
\usepackage{tcolorbox}
\usepackage[T1]{fontenc}
\usepackage[english]{babel}
\usepackage{amsthm}
\usepackage{amsfonts}
\usepackage[ruled,vlined,linesnumbered]{algorithm2e}
\usepackage{amssymb}
\usepackage{pifont}
\usepackage{makecell}
\usepackage{fontawesome}

\definecolor{Gray}{gray}{0.85}
\newcommand\crule[3][Gray]{\textcolor{#1}{\rule{#2}{#3}}}

\theoremstyle{definition}
\newtheorem{definition}{Definition}[section]

\input{languages}

\newcommand*\wrapletters[1]{\wr@pletters#1\@nil}
\def\wr@pletters#1#2\@nil{#1\allowbreak\if&#2&\else\wr@pletters#2\@nil\fi}

\newcommand{\block}[1]{\href{https://etherscan.io/block/#1}{#1}\xspace}
\newcommand{\etherscanTx}[1]{\href{https://etherscan.io/tx/#1}{\wrapletters{#1}}\xspace}
\newcommand{\abbrEtherscanTx}[1]{\href{https://etherscan.io/tx/#1}{\StrLeft{#1}{6}..\StrRight{#1}{4}}\xspace}
\newcommand{\etherscanAddress}[1]{\href{https://etherscan.io/address/#1}{\wrapletters{#1}}\xspace}

\newcommand{\bscblock}[1]{\href{https://bscscan.com/block/#1}{#1}\xspace}
\newcommand{\bscscanTx}[1]{\href{https://bscscan.com/tx/#1}{\wrapletters{#1}}\xspace}

\newcommand{\bscscanAddress}[1]{\href{https://bscscan.com/address/#1}{\wrapletters{#1}}\xspace}

\newcommand{\etal}{\textit{et al.}}
\npthousandsep{,}
\npdecimalsign{.}

\usepackage{acro}
\acsetup{single}

\DeclareAcronym{DeFi}{
  short = DeFi,
  long  = Decentralized Finance,
}
\newcommand{\DeFi}{\ac{DeFi}\xspace}

\DeclareAcronym{CeFi}{
  short = CeFi,
  long  = Centralized Finance,
}
\DeclareAcronym{DApps}{
  short = dApps,
  long  = Decentralized Applications,
}

\DeclareAcronym{MEV}{
  short = MEV,
  long  = Miner Extractable Value,
}
\newcommand{\MEV}{\ac{MEV}\xspace}

\DeclareAcronym{BEV}{
  short = BEV,
  long  = Blockchain Extractable Value,
}
\newcommand{\BEV}{\ac{BEV}\xspace}

\DeclareAcronym{CFG}{
  short = CFG,
  long  = control-flow graph,
}
\newcommand{\CFG}{\ac{CFG}\xspace}

\DeclareAcronym{DCFG}{
  short = DCFG,
  long  = dynamic control-flow graph,
}
\newcommand{\DCFG}{\ac{DCFG}\xspace}

\DeclareAcronym{FaaS}{
  short = FaaS,
  long  = Front-running as a Service,
}
\newcommand{\FaaS}{\ac{FaaS}\xspace}

\DeclareAcronym{P2P}{
  short = P2P,
  long  = peer-to-peer,
}
\newcommand{\PtoP}{\ac{P2P}\xspace}

\DeclareAcronym{DApp}{
  short = DApp,
  long  = Decentralized Application,
}
\newcommand{\DApp}{\ac{DApp}\xspace}

\DeclareAcronym{IPS}{
  short = IPS,
  long  = Intrusion Prevention System,
}

\DeclareAcronym{EVM}{
  short = EVM,
  long  = Ethereum Virtual Machine,
}
\newcommand{\EVM}{\ac{EVM}\xspace}

\DeclareAcronym{BSC}{
  short = BSC,
  long  = BNB Smart Chain,
}
\newcommand{\BSC}{\ac{BSC}\xspace}

\DeclareAcronym{ABI}{
  short = ABI,
  long  = Application Binary Interface,
}
\newcommand{\ABI}{\ac{ABI}\xspace}

\DeclareAcronym{AMM}{
  short = AMM,
  long  = automated market maker,
}

\newcommand{\adversary}{$\mathcal{A}$\xspace}
\newcommand{\txv}{$\mathsf{tx}_v$\xspace}
\newcommand{\txc}{$\mathsf{tx}_c$\xspace}
\newcommand{\scvi}{$\mathsf{sc}_{v_i}$\xspace}
\newcommand{\scvj}{$\mathsf{sc}_{v_j}$\xspace}

\newcommand{\tscvi}{$\widehat{\mathsf{sc}_{v_i}}$\xspace}

\newcommand{\rscvi}{$\overline{\mathsf{sc}_{v_i}}$\xspace}
\newcommand{\scvis}{$\{\mathsf{sc}_{v_i}\}$\xspace}
\newcommand{\scai}{$\mathsf{sc}_{a_i}$\xspace}
\newcommand{\scaj}{$\mathsf{sc}_{a_j}$\xspace}
\newcommand{\scais}{$\{\mathsf{sc}_{a_i}\}$\xspace}

\newcommand{\JUMPI}{\texttt{JUMPI}\xspace}
\newcommand{\JUMP}{\texttt{JUMP}\xspace}

\newcommand{\empirical}[1]{#1}

\newcommand{\StartDate}{\empirical{1st of August,~2021}\xspace}
\newcommand{\EndDate}{\empirical{31st of July,~2022}\xspace}
\newcommand{\TimeDuration}{\empirical{one year}\xspace}

\newcommand{\ETHStartBlock}{\empirical{\block{12936340}}\xspace}
\newcommand{\ETHEndBlock}{\empirical{\block{15253305}}\xspace}

\newcommand{\UpfrontLessThanFiveEtherPercentage}{\empirical{$99.31\%$}\xspace}

\newcommand{\potentialVictimTransactions}{\empirical{$\numprint{431416565}$}\xspace}
\newcommand{\naiveReplayableTransactions}{\empirical{$\numprint{43979}$}\xspace}
\newcommand{\naiveReplayablePercentage}{\empirical{$0.0102\%$}\xspace}
\newcommand{\newAttackTransactions}{\empirical{$\numprint{26127}$}\xspace}
\newcommand{\newAttackTransactionPercentage}{\empirical{$0.0061\%$}\xspace}
\newcommand{\newAttackContracts}{\empirical{$\numprint{665}$}\xspace}

\newcommand{\averageContractToReplace}{\empirical{$1.02\pm0.15$}\xspace}
\newcommand{\averageContractToReplaceMax}{\empirical{$3$}\xspace}
\newcommand{\averageContractToReplaceMin}{\empirical{$1$}\xspace}
\newcommand{\averageContractToReplaceMean}{\empirical{$1.02$}\xspace}
\newcommand{\averageContractToReplaceStd}{\empirical{$0.15$}\xspace}

\newcommand{\contractSizeReduction}{\empirical{$60.95\pm19.19\%$}\xspace}
\newcommand{\contractSizeReductionMax}{\empirical{$98.63\%$}\xspace}
\newcommand{\contractSizeReductionMin}{\empirical{$-295.56\%$}\xspace}
\newcommand{\contractSizeReductionMinNegative}{\empirical{$295.56\%$}\xspace}
\newcommand{\contractSizeReductionMean}{\empirical{$60.16\%$}\xspace}
\newcommand{\contractSizeReductionStd}{\empirical{$19.19\%$}\xspace}

\newcommand{\naiveReplayProfit}{\empirical{$\numprint{13.87}$M~USD}\xspace}
\newcommand{\naiveReplayProfitNoUnit}{\empirical{$\numprint{13.87}$M}\xspace}

\newcommand{\newAttackProfit}{\empirical{$\numprint{148.96}$M~USD}\xspace}

\newcommand{\newAttackAdditionalProfit}{\empirical{$\numprint{135.08}$M~USD}\xspace}
\newcommand{\newAttackAdditionalProfitNoUnit}{\empirical{$\numprint{135.08}$M}\xspace}

\newcommand{\naiveReplayProfitAverageNoUnit}{\empirical{$315.48\pm4.73$K}\xspace}

\newcommand{\newAttackProfitAverageNoUnit}{\empirical{$5.17$K$\pm227.22$K}\xspace}

\newcommand{\profitIncreaseRatio}{\empirical{$\numprint{973.6}\%$}\xspace}

\newcommand{\naiveReplayGasConsumption}{\empirical{$0.45$M$\pm0.50$M}\xspace}
\newcommand{\newAttackGasConsumption}{\empirical{$0.98$M$\pm0.74$M}\xspace}
\newcommand{\newAttackGasConsumptionMax}{\empirical{$23.11$M}\xspace}
\newcommand{\newAttackGasContractDeploymentRatio}{\empirical{$48.42\%$}\xspace}

\newcommand{\blockLimitAverage}{\empirical{$29.77$M}\xspace}

\newcommand{\AnalyzedVictims}{\empirical{$100$}\xspace}
\newcommand{\ArbLiqTransactions}{\empirical{$55$}\xspace}
\newcommand{\ArbLiqProfitNoUSD}{\empirical{$8.66$M}\xspace}

\newcommand{\AnalyzedVictimTotalProfit}{\empirical{$113.43$M~USD}\xspace}
\newcommand{\AnalyzedVictimTotalProfitPercentage}{\empirical{$83.97\%$}\xspace}

\newcommand{\DeFiAttacks}{\empirical{$13$}\xspace}
\newcommand{\DeFiAttackTotalTransactions}{\empirical{$29$}\xspace}

\newcommand{\DeFiAttackBlackhatTransactions}{\empirical{$17$}\xspace}
\newcommand{\DeFiAttackWhitehatTransactions}{\empirical{$12$}\xspace}
\newcommand{\DeFiAttackWhitehatTransactionsPrivate}{\empirical{$10$}\xspace}

\newcommand{\PopsicleProfit}{\empirical{$20.25$M}\xspace}

\newcommand{\DeFiAttackProfit}{\empirical{$73.74$M~USD}\xspace}
\newcommand{\DeFiAttackProfitNoUSD}{\empirical{$73.74$M}\xspace}

\newcommand{\MassDepositLoss}{\empirical{$28.58$M~USD}\xspace}

\newcommand{\newVulTransactions}{\empirical{$14$}\xspace}
\newcommand{\newVulProfitNoUSD}{\empirical{$30.23$M}\xspace}

\newcommand{\knownVulTransactions}{\empirical{$2$}\xspace}
\newcommand{\knownVulProfitNoUSD}{\empirical{$795.86$K}\xspace}

\newcommand{\ETHRealtimeEffectiveTime}{\empirical{$26$~days}\xspace}
\newcommand{\ETHIncomingTxRate}{\empirical{$17.86$}\xspace}

\newcommand{\ETHRealtimeTxBoth}{\empirical{$\numprint{4045}$}\xspace}

\newcommand{\ETHRealtimeTxNewAttack}{\empirical{$\numprint{3699}$}\xspace}

\newcommand{\naiveTimeConsumption}{\empirical{$0.01\pm0.01$}\xspace}

\newcommand{\naiveTimeConsumptionMean}{\empirical{$0.01$}\xspace}
\newcommand{\naiveTimeConsumptionStd}{\empirical{$0.01$}\xspace}
\newcommand{\naiveTimeConsumptionMax}{\empirical{$0.11$}\xspace}
\newcommand{\naiveTimeConsumptionMin}{\empirical{$2\times10^{-4}$}\xspace}

\newcommand{\timeConsumption}{\empirical{$0.07\pm0.10$}\xspace}
\newcommand{\StepOneThreeSixPercentage}{\empirical{$96.35\%$}\xspace}

\newcommand{\timeConsumptionMean}{\empirical{$0.07$}\xspace}
\newcommand{\timeConsumptionStd}{\empirical{$0.10$}\xspace}
\newcommand{\timeConsumptionMax}{\empirical{$1.59$}\xspace}
\newcommand{\timeConsumptionMin}{\empirical{$9\times10^{-4}$}\xspace}

\newcommand{\steponetimemean}{\empirical{$0.02$}\xspace}
\newcommand{\steponetimestd}{\empirical{$0.03$}\xspace}
\newcommand{\steponetimemax}{\empirical{$0.36$}\xspace}
\newcommand{\steponetimemin}{\empirical{$3\times10^{-4}$}\xspace}

\newcommand{\steptwotimemean}{\empirical{$2\times10^{-3}$}\xspace}
\newcommand{\steptwotimestd}{\empirical{$5\times10^{-3}$}\xspace}
\newcommand{\steptwotimemax}{\empirical{$0.10$}\xspace}
\newcommand{\steptwotimemin}{\empirical{$2\times10^{-5}$}\xspace}

\newcommand{\stepthreetimemean}{\empirical{$0.04$}\xspace}
\newcommand{\stepthreetimestd}{\empirical{$0.06$}\xspace}
\newcommand{\stepthreetimemax}{\empirical{$1.39$}\xspace}
\newcommand{\stepthreetimemin}{\empirical{$3\times10^{-4}$}\xspace}

\newcommand{\stepfourtimemean}{\empirical{$2\times10^{-5}$}\xspace}
\newcommand{\stepfourtimestd}{\empirical{$5\times10^{-5}$}\xspace}
\newcommand{\stepfourtimemax}{\empirical{$2\times10^{-3}$}\xspace}
\newcommand{\stepfourtimemin}{\empirical{$1\times10^{-6}$}\xspace}

\newcommand{\stepfivetimemean}{\empirical{$5\times10^{-4}$}\xspace}
\newcommand{\stepfivetimestd}{\empirical{$2\times10^{-3}$}\xspace}
\newcommand{\stepfivetimemax}{\empirical{$0.09$}\xspace}
\newcommand{\stepfivetimemin}{\empirical{$2\times10^{-5}$}\xspace}

\newcommand{\stepsixtimemean}{\empirical{$7\times10^{-3}$}\xspace}
\newcommand{\stepsixtimestd}{\empirical{$0.02$}\xspace}
\newcommand{\stepsixtimemax}{\empirical{$0.96$}\xspace}
\newcommand{\stepsixtimemin}{\empirical{$2\times10^{-4}$}\xspace}

\newcommand{\ETHRealtimeSuccessRate}{\empirical{$99.68\%$}\xspace}
\newcommand{\ETHRealtimeMempoolPerformance}{\empirical{$12.56\pm12.55$}\xspace}

\newcommand{\BSCStartBlock}{\empirical{\bscblock{9643812}}\xspace}
\newcommand{\BSCEndBlock}{\empirical{\bscblock{20045094}}\xspace}

\newcommand{\UpfrontLessThanFiveBNBPercentage}{\empirical{$99.48\%$}\xspace}

\newcommand{\BSCpotentialVictimTransactions}{\empirical{$\numprint{2366970381}$}\xspace}
\newcommand{\BSCnaiveReplayableTransactions}{\empirical{$\numprint{516128}$}\xspace}
\newcommand{\BSCnaiveReplayablePercentage}{\empirical{$0.0218\%$}\xspace}
\newcommand{\BSCnewAttackTransactions}{\empirical{$\numprint{52799}$}\xspace}
\newcommand{\BSCnewAttackTransactionPercentage}{\empirical{$0.0022\%$}\xspace}
\newcommand{\BSCnewAttackContracts}{\empirical{$\numprint{1193}$}\xspace}

\newcommand{\BSCaverageContractToReplace}{\empirical{$1.05\pm0.23$}\xspace}
\newcommand{\BSCaverageContractToReplaceMax}{\empirical{$3$}\xspace}
\newcommand{\BSCaverageContractToReplaceMin}{\empirical{$1$}\xspace}
\newcommand{\BSCaverageContractToReplaceMean}{\empirical{$1.05$}\xspace}
\newcommand{\BSCaverageContractToReplaceStd}{\empirical{$0.23$}\xspace}

\newcommand{\BSCcontractSizeReduction}{\empirical{$57.59\pm18.69\%$}\xspace}
\newcommand{\BSCcontractSizeReductionMax}{\empirical{$99.46\%$}\xspace}
\newcommand{\BSCcontractSizeReductionMin}{\empirical{$-613.33\%$}\xspace}
\newcommand{\BSCcontractSizeReductionMinNegative}{\empirical{$613.33\%$}\xspace}
\newcommand{\BSCcontractSizeReductionMean}{\empirical{$57.59\%$}\xspace}
\newcommand{\BSCcontractSizeReductionStd}{\empirical{$18.69\%$}\xspace}

\newcommand{\BSCnaiveReplayProfit}{\empirical{$\numprint{13.25}$M~USD}\xspace}
\newcommand{\BSCnaiveReplayProfitNoUnit}{\empirical{$\numprint{13.25}$M}\xspace}

\newcommand{\BSCnewAttackProfit}{\empirical{$\numprint{42.70}$M~USD}\xspace}

\newcommand{\BSCnewAttackAdditionalProfit}{\empirical{$\numprint{29.45}$M~USD}\xspace}
\newcommand{\BSCnewAttackAdditionalProfitNoUnit}{\empirical{$\numprint{29.45}$M}\xspace}

\newcommand{\BSCnaiveReplayProfitAverageNoUnit}{\empirical{$25.67\pm1.78$K}\xspace}

\newcommand{\BSCnewAttackProfitAverageNoUnit}{\empirical{$557.75\pm55.88$K}\xspace}

\newcommand{\BSCprofitIncreaseRatio}{\empirical{$\numprint{222.3}\%$}\xspace}

\newcommand{\BSCnaiveReplayGasConsumption}{\empirical{$1.74$M$\pm4.44$B}\xspace}
\newcommand{\BSCnewAttackGasConsumption}{\empirical{$1.64$M$\pm0.64$B}\xspace}

\newcommand{\BSCnewAttackGasContractDeploymentRatio}{\empirical{$53.04\%$}\xspace}

\newcommand{\BSCblockLimitAverage}{\empirical{$82.66$M}\xspace}

\newcommand{\BSCGasOverThreeM}{\empirical{$\numprint{82192}$}\xspace}
\newcommand{\BSCGasToken}{\empirical{$\numprint{80291}$}\xspace}
\newcommand{\BSCnaiveReplayGasConsumptionWOGasToken}{\empirical{$0.48$M$\pm0.55$M}\xspace}

\newcommand{\BSCArbLiqTransactions}{\empirical{$13$}\xspace}
\newcommand{\BSCArbLiqProfitNoUSD}{\empirical{$506.07$K}\xspace}

\newcommand{\BSCAnalyzedVictimTotalProfit}{\empirical{$27.27$M~USD}\xspace}
\newcommand{\BSCAnalyzedVictimTotalProfitPercentage}{\empirical{$92.60\%$}\xspace}

\newcommand{\BSCDeFiAttacks}{\empirical{$22$}\xspace}
\newcommand{\BSCDeFiAttackTotalTransactions}{\empirical{$40$}\xspace}

\newcommand{\BSCDeFiAttackProfit}{\empirical{$22.39$M~USD}\xspace}
\newcommand{\BSCDeFiAttackProfitNoUSD}{\empirical{$22.39$M}\xspace}
\newcommand{\BSCMassDepositLoss}{\empirical{$759.54$K~USD}\xspace}

\newcommand{\BSCNewVulTransactions}{\empirical{$16$}\xspace}
\newcommand{\BSCNewVulProfitNoUSD}{\empirical{$1.30$M}\xspace}

\newcommand{\BSCRealtimeEffectiveTime}{\empirical{$14$~days}\xspace}
\newcommand{\BSCIncomingTxRate}{\empirical{$57.31$}\xspace}

\newcommand{\BSCRealtimeTxNaive}{\empirical{$\numprint{784}$}\xspace}
\newcommand{\BSCRealtimeTxNewAttack}{\empirical{$\numprint{489}$}\xspace}

\newcommand{\BSCRealtimeSuccessRate}{\empirical{$99.49\%$}\xspace}
\newcommand{\BSCRealtimeMempoolPerformance}{\empirical{$2.24\pm0.81$}\xspace}

\newcommand{\knownAttacksEthAndBsc}{\empirical{$35$}\xspace}
\newcommand{\newVulProfitEthAndBsc}{\empirical{$31.53$M}\xspace}

\newcommand{\TotalReplayProfitUSD}{\empirical{$35.37$M~USD}\xspace}

\newcommand{\attackname}{\textsc{Ape}\xspace}

\begin{document}


\title{The Blockchain Imitation Game}

\author{
{\rm Kaihua Qin}\\
Imperial College London, RDI
\and
{\rm Stefanos Chaliasos}\\
Imperial College London
\and
{\rm Liyi Zhou}\\
Imperial College London, RDI
\and
{\rm Benjamin Livshits}\\
Imperial College London
\and
{\rm Dawn Song}\\
UC Berkeley, RDI
\and
{\rm Arthur Gervais}\\
University College London, RDI
} 


\maketitle


\begin{abstract}
The use of blockchains for automated and adversarial trading has become commonplace. However, due to the transparent nature of blockchains, an adversary is able to observe any pending, not-yet-mined transactions, along with their execution logic. This transparency further enables a new type of adversary, which copies and front-runs profitable pending transactions in real-time, yielding significant financial gains.

Shedding light on such ``copy-paste'' malpractice, this paper introduces the Blockchain Imitation Game and proposes a generalized imitation attack methodology called \attackname. Leveraging dynamic program analysis techniques, \attackname supports the automatic synthesis of adversarial smart contracts. Over a timeframe of \TimeDuration (\StartDate to \EndDate), \attackname could have yielded~\newAttackProfit in profit on Ethereum, and~\BSCnewAttackProfit on \BSC.

Not only as a malicious attack, we further show the potential of transaction and contract imitation as a defensive strategy. Within \TimeDuration, we find that \attackname could have successfully imitated \DeFiAttacks and \BSCDeFiAttacks known \DeFi attacks on Ethereum and \BSC, respectively. Our findings suggest that blockchain validators can imitate attacks in real-time to prevent intrusions in \DeFi.

\end{abstract}

\acresetall

\section{Introduction}
\DeFi, built upon blockchains, has grown to a multi-billion USD industry. However, blockchain \PtoP networks have been described as dark forests, where traders engage in competitive trading, indulging in adversarial front-running~\cite{daian2020flash}. Such front-running is possible, because of the inherent time delay between a transaction's creation, and its being committed on the blockchain. This time delay often lasts only a few seconds, posing computational challenges for the front-running players. To yield a financial revenue, a \DeFi trader needs to monitor the convoluted market dynamics and craft profitable transactions promptly, which typically requires professional domain knowledge. Alternatively, an adversarial trader may also seek to ``copy-paste'' and front-run a pending profitable transaction without understanding its logic. We term such a strategy as an \emph{imitation attack}. A naive string-replace imitation method~\cite{qin2022quantifying} was shown to yield thousands of USD per month on past blockchain states. The practitioners' community swiftly came up with defenses to counter such a naive imitation method. At the time of writing, traders often deploy personalized and closed-source smart contracts, making exploitation harder. The known naive imitation algorithm no longer applies, because these contracts are typically protected through, for example, authentications.

\begin{figure}[tb]
    \centering
    \includegraphics[width=\columnwidth]{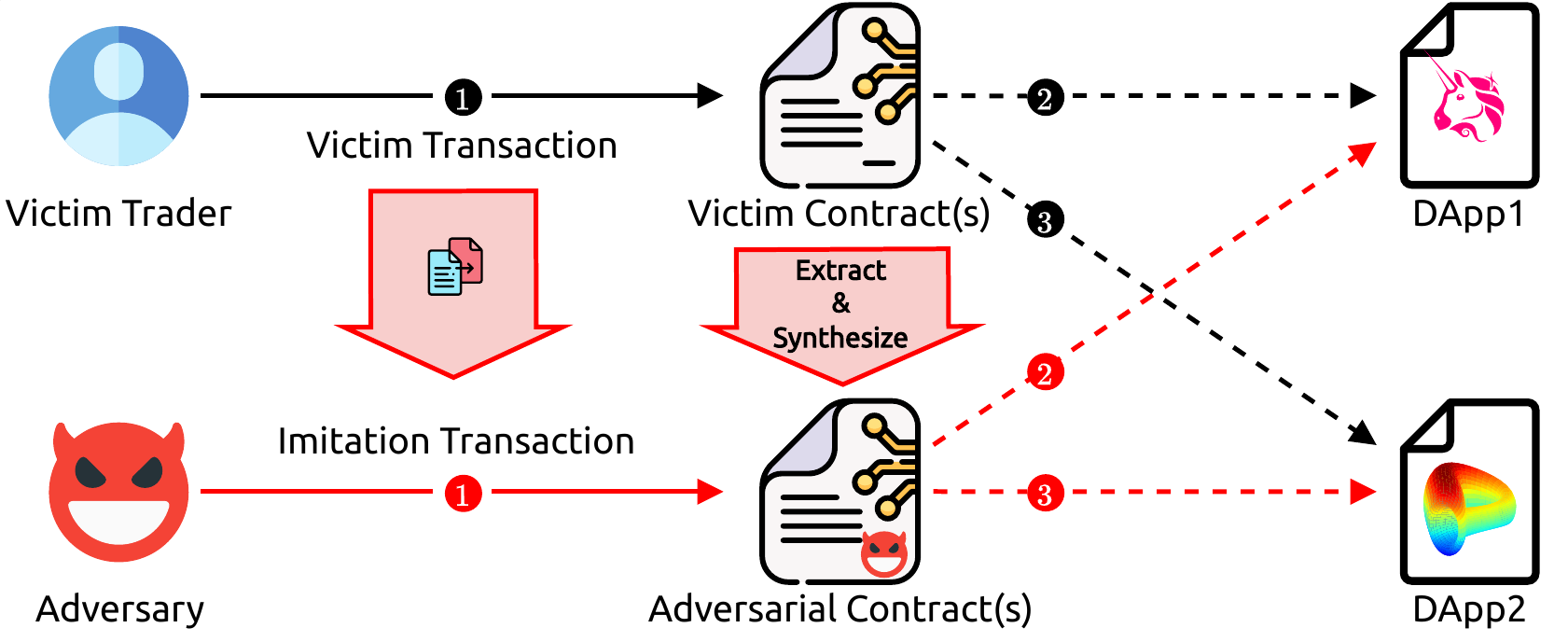}
    \caption{High-level \attackname attack mechanism, a generalized, automated imitation method synthesizing adversarial contracts without prior knowledge about the victim's transaction and contract(s). \attackname appropriates any resulting revenue.}
    \label{fig:introductory}
\end{figure}





However, the possibility of a generalized imitation attack that can invalidate existing protection mechanisms has not yet been explored. The goal of this work is to investigate, design, implement, and evaluate a generalized imitation method.
We find that, to successfully imitate a transaction, an attacker needs to overcome the following three technical challenges.
\textbf{(I)} The short front-running time-window may exclude the application of powerful program analysis techniques, such as symbolic executions, which are not designed for real-time tasks.
\textbf{(II)} An attacker needs to recursively identify the victim contracts that hinder the imitation execution, and replace them with newly synthesized adversarial contracts. Blockchain virtual machine instrumentation is hence necessary to ensure the efficiency of this identification process.
\textbf{(III)} An attacker must guarantee that the synthesized contracts are invoked and executed correctly, while the generated financial revenue is sent to an adversarial account after the imitation execution. To the best of our knowledge, no existing work nor off-the-shelf tool allows automatically copying and synthesizing smart contracts with custom logic injected. 

In this work, we propose \attackname (cf.\ Figure~\ref{fig:introductory}), an automated generalized imitation methodology. On a high level, \attackname is designed to construct an imitation transaction that copies the logic of a victim transaction. When it is necessary, \attackname synthesizes and deploys adversarial smart contracts to bypass copy protections, such as authentication mechanisms. To this end, \attackname leverages dynamic taint analysis, program synthesis, and advanced instrumentations to realize imitation generation. The \emph{generalization} stems from the fact that the adversary does not require prior victim knowledge, nor needs to understand the victim transaction logic, nor requires prior knowledge of the interacting \DApp. \attackname applies to e.g., trading activities for fungible, non-fungible tokens, and exploit transactions.

Despite the blockchain-based \DeFi domain flourishing, it is plagued by multi-million dollar hacks~\cite{zhou2022sok}. As outlined in this work, an imitation attack can yield a significant financial profit to an adversary. However, following a defence-in-depth approach, a blockchain imitation can also act as an intrusion prevention system by mimicking an attack, appropriating the vulnerable funds, and returning them to its victim. The practitioner community has dubbed such benevolent activity as ``whitehat hacking''.

We summarize our contributions as follows.
\begin{itemize}
    \item We introduce the generalized blockchain imitation game with a new class of adversary attempting to imitate its victim transactions and associated contracts, without prior knowledge about the victim's intent or application logic. We design \attackname, a generalized imitation tool for EVM-based blockchains. We are the first to show that dynamic program analysis techniques can realize an imitation attack, posing a substantial threat to blockchain users.
    
    \item We evaluate~\attackname over a~\TimeDuration timeframe on Ethereum and \BSC. 
    We show that \attackname could have yielded~\newAttackProfit in profit on Ethereum, and~\BSCnewAttackProfit on \BSC.
    We find that~\DeFiAttackProfit stems from~\knownAttacksEthAndBsc known \DeFi attacks that \attackname can imitate. \attackname's impact further becomes apparent through the discovery of five new vulnerabilities, which we responsibly disclose, as they could have caused a total loss of~\newVulProfitEthAndBsc~USD, if exploited.

    \item We show that \attackname executes in real-time on Ethereum ($13.3$-second inter-block time) and \BSC ($3$-second inter-block time). On average, a single \attackname imitation takes~\timeConsumption seconds. Because of \attackname's efficiency, it could have front-run in real-time~\knownAttacksEthAndBsc \DeFi attacks within our evaluation timeframe. Miners that execute \attackname can ultimately choose to carry out the attacks, or could act as whitehat hackers in a defensive capacity.


\end{itemize}

\section{Background}
In this section, we outline the required background and provide motivating examples.

\subsection{Blockchain and Smart Contract}
A blockchain is a chain of blocks distributed over a \PtoP network~\cite{bitcoin}. Users approve transactions through public-key signatures from an account. Miners collect transactions and mine those in a specific sequence within blocks. Smart contract-enabled blockchains extend the capabilities of accounts to hold assets and code, which can perform arbitrary computations. Smart contracts are initialized and executed through transactions and remain immutable once deployed. They are usually written in high-level languages (e.g., Solidity) that are compiled to low-level bytecode, executed by a blockchain's virtual machine, e.g., \EVM~\cite{wood2014ethereum}. The \EVM is a stack-based virtual machine supporting arithmetic, control-flow, cryptographic and other blockchain-specific instructions (e.g., accessing the current block number). Each executed bytecode instruction costs gas, paid with the native cryptocurrency. Note that all code and account balances are typically transparently visible. The \ABI of a smart contract defines how a smart contract function should be invoked (including the function type, name, input parameters, etc). If a smart contract is not open-source, the \ABI is likely unavailable.

\DeFi~\cite{qin2021cefi}, an emerging financial ecosystem built on top of smart contract blockchains, at the time of writing, reached a peak of~$300$B USD total value locked.\footnote{\url{https://defillama.com/}} At its core, \DeFi is a smart contract encoded financial ecosystem, implementing, e.g., automated market maker~\cite{angeris2020improved}, lending platforms~\cite{qin2021empirical,wang2022speculative}, and stablecoins~\cite{moin2020sok,clark2019sok}. Users can access a \DeFi application (a \DApp) by issuing transactions to the respective contracts.

\subsection{Blockchain Extractable Value}
It is well-known that Wall Street traders profit by front-running other investors' orders (i.e., high-frequency trading)~\cite{lewis2014flash}. Similarly, the transaction order in \DeFi fatefully impacts the revenue extraction activities. By default, miners order transactions on a descending transaction fee basis. Therefore, \DeFi traders can front- and back-run pending (i.e., not yet mined) transactions by competitively offering a transaction fee~\cite{eskandari2019sok,qin2022quantifying}. 
Miners, however, have the single-handed privilege to order transactions, which grants them a monopoly on blockchain value extraction. This privilege leads to the concept of \MEV~\cite{daian2020flash}. Qin \etal~\cite{qin2022quantifying} generalize \MEV to \BEV and show that over $32$ months, the extracted \BEV on Ethereum amounts to~$\numprint{540.54}$M USD, contributed by three major sources, sandwich attacks~\cite{zhou2021high}, liquidations~\cite{qin2021empirical,qin2023mitigating}, and arbitrage~\cite{zhou2021just}. Note that \FaaS (e.g., \href{https://docs.flashbots.net/}{flashbots}) reduces the risks of extracting \BEV, by colluding with miners in a private network. Similar to bribes~\cite{bonneau2016buy}, \BEV is proven to threaten the blockchain consensus security because miners are incentivized to fork the blockchain~\cite{daian2020flash,zhou2021just,qin2022quantifying}. An SoK on \DeFi attacks systematizes attacks over a timeframe of four years~\cite{zhou2022sok}, identifying the various attack causes and implications. In this paper we make the distinction among whitehat and blackhat attackers, wherein a blackhat attacker is one who retains the financial proceeds of an attack, whereas a whitehat refunds the attack revenue to the identified victim.



\subsection{Naive Transaction Imitation Attack}\label{sec:transaction-replay-attack}
\BEV extracting entities can follow two strategies: either meticulously analyzing \DeFi applications (i.e., application-specific extraction), or applying a naive generalized transaction imitation attack (i.e., application-agnostic extraction).

Qin \etal~\cite{qin2022quantifying} propose a naive but effective transaction imitation attack. This algorithm takes as input a victim transaction, and simply replaces the transaction's sender address with an adversarial address in the transaction sender and data fields. As such, this imitation algorithm corresponds to a string replacing approach, and \emph{does not attempt to synthesize adversarial smart contracts}. When simulating on past blockchain data, it was shown that such naive algorithm could have generated~\TotalReplayProfitUSD over a~$32$ month timeframe. We provide a naive imitation example in Appendix~\ref{app:naive-replay-attack-example}. This algorithm, however, fails when traders protect their transaction through, for example, authentication mechanisms as we outline below.

\subsection{Motivating Examples}\label{sec:motivating-example}
\begin{figure}[tb]
\centering
\includegraphics[width=0.8\columnwidth]{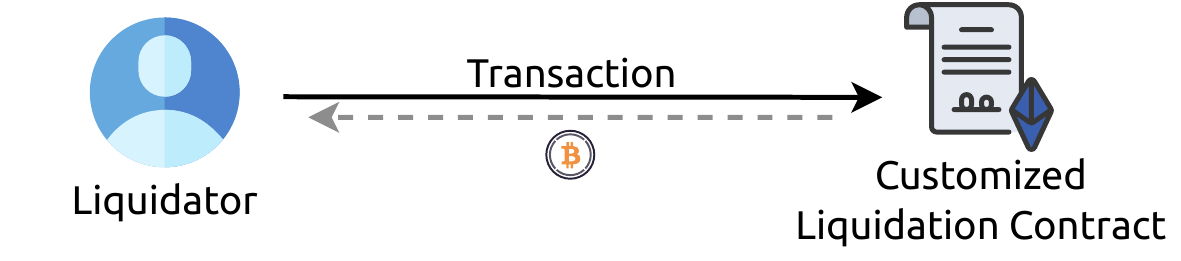}

\lstset{language=Decompiled}
\lstinputlisting[lastline=2,frame=lrt,belowskip=0pt]{decompiled}
\lstinputlisting[backgroundcolor=\color{lightgray},firstline=3,firstnumber=3,lastline=3,frame=lr,aboveskip=0pt,belowskip=0pt]{decompiled}
\lstinputlisting[firstline=4,firstnumber=4,frame=lrb,aboveskip=0pt]{decompiled}

\caption{Motivating example. A liquidator triggers a liquidation by sending a transaction to its customized contract. The customized contract (cf.\ \etherscanAddress{0xe0a9efE32985cC306255b395a1bd06D21ccEAd42}) contains a authentication (line $3$).}
\label{fig:non-replayable-transaction}
\end{figure}

Traders in \DeFi typically deploy customized smart contracts to perform financial actions, such as arbitrage and liquidations.\footnote{A liquidation refers to the process of selling collateral to secure debt~\cite{qin2021empirical}.} Those traders should protect their transactions from imitation attacks, by for instance only allowing predefined accounts to invoke their smart contract (cf.\ Figure~\ref{fig:non-replayable-transaction}). Such protection corresponds to an authentication mechanism, a common practice that the practitioners' community employs. We proceed with a real-world authentication example.


\emph{Authentication Protection Example.} In Figure~\ref{fig:non-replayable-transaction}, we show how a liquidation contract\footnote{Address: \etherscanAddress{0x18C0cA3947E255881f94DE50B3a906Fc2759F7FE}.} attempts to prevent a transaction imitation attack. The solidity code of Figure~\ref{fig:non-replayable-transaction}'s liquidation contract is not open-source, and we therefore decompile the contract bytecode with a state-of-the-art \EVM contract decompiler~\cite{neville2022elipmoc}. To trigger a liquidation, the liquidator needs to send a transaction\footnote{E.g., \etherscanTx{0x631a4941eb8d0903c1c0073784423f87019cddd7c3822c77258172bd8d1a862c}.} to the augmented contract along with parameters (e.g., the flash loan size and liquidation amount). If there is no protection mechanism, an adversary might front-run the liquidator, by calling the same contract with identical parameters. However, the presented contract requires the transaction sender to match a specific address (in line $3$). Specifically, the liquidation function \texttt{printMoney()} is only callable by a hard-coded address. If this condition is not met, the naive imitation attack from Section~\ref{sec:transaction-replay-attack} fails. To circumvent authentication protections, one approach is to synthesize an adversarial contract that replicates the liquidator's customized contract. By reconstructing the bytecode, the synthesized contract preserves the liquidation logic while bypassing the authentication. To perform the attack, the adversary does not need to understand the business logic of the liquidation transaction. Instead of invoking the liquidator's customized contract, the adversary invokes the synthesized contract, which then triggers a liquidation.

In this work, we present an automated and generalized imitation methodology, \attackname, that thwarts such protections. Not only limited to authentications, \attackname provides a comprehensive approach to overcome various forms of protection mechanisms that cannot be handled by the naive imitation strategy. As another example, we consider a scenario where a trader issues a profitable transaction that deposits the earned revenue in a smart contract under the trader's control. While an adversary may succeed in executing an imitation transaction through the naive strategy, the revenue would remain in the ``victim'' trader's contract, resulting in no financial gain for the attacker. In contrast, \attackname resolves this predicament by synthesizing an adversarial contract that serves as the revenue recipient. Further details of this scenario are presented in Section~\ref{sec:analysis}, where we showcase a real-world transaction.

\section{\attackname Overview}
We proceed to outline the system and threat model. We then overview the key components of \attackname.

\subsection{Preliminary Models}\label{sec:models}
\paragraph{System Model} We consider a smart-contract-enabled distributed ledger with an existing \DeFi ecosystem. A trader performs financial actions through its blockchain account, signing transactions mined by miners. Similarly, smart contracts are referenced by their respective account. A ledger is a state machine replication~\cite{schneider1990implementing}, with state $S$. A blockchain transaction $\mathsf{tx}$ represents a state transition function, converting the ledger state from $S$ to $S'$, i.e., $S'=\mathsf{tx}(S)$.

We assume that, within the \DeFi system, there exist various asset representations (e.g., fungible tokens) in addition to the native blockchain cryptocurrency $E$. The balance of asset $m$ held by an account $a$ at the blockchain state $S$ is denoted by $\operatorname{Bal}^m_a(S)$, while $\Delta^m_a(S, \mathsf{tx})$ denotes the balance change after executing a transaction $\mathsf{tx}$ upon $S$ (cf.\ Equation~\ref{eq:balance-delta}).
\begin{equation}\label{eq:balance-delta}
    \Delta^m_a(S, \mathsf{tx}) = \operatorname{Bal}^m_a\left(\mathsf{tx}(S)\right) - \operatorname{Bal}^m_a(S) 
\end{equation}
We further assume the existence of on-chain exchanges which allow trades from any asset $m$ to the native cryptocurrency $E$.



\begin{table}[t]
    \centering
    \caption{Notations adopted in this work.}
    \resizebox{\columnwidth}{!}{%
    \begin{tabular}{ll}
    \toprule
    Symbol    &  Description  \\\midrule
    $S$   & A blockchain state \\
    $\mathsf{tx}_v$ & Victim transaction \\
    \scvi & Victim contract $i$ \\
    \tscvi & Tainted contract $i$ \\
    \rscvi & Contract $i$ to be patched and replaced \\
    \txc & Adversarial imitation transaction \\
    \scai & Adversarial synthesized contract $i$ \\
    $E$    & The native cryptocurrency \\
    $\operatorname{Bal}^m_a(S)$    & Balance of asset $m$ held by account $a$ \\
    $\Delta^m_a(S, \mathsf{tx})$    & Balance change of $m$ held by $a$ after executing $\mathsf{tx}$ upon $S$\\
    \bottomrule
    \end{tabular}}
    \label{tab:notations}
\end{table}

\paragraph{Threat Model}\label{sec:threatmodel}
Our strongest possible adversarial model considers a miner that can single-handedly order transactions in its mined blocks. 
As a miner, \adversary has access to every pending transaction and the current blockchain state. \adversary actively listens for data on the P2P network through multiple distributed nodes and peer connections. \adversary can moreover act as, or collaborate with a \FaaS provider to privately receive pending transactions. Because \adversary is a miner, \adversary can front-run any pending transaction. We assume that \adversary is financially rational and attempts to maximize its asset value. Moreover, we assume that \adversary has access to sufficient $E$ to execute \attackname. 

\attackname is application-agnostic, meaning that \adversary does not need to have any upfront knowledge of the logic of a victim transaction \txv, nor its target smart contract. We, however, assume that \adversary understands how to interact with asset tokens and that \adversary can swap a token $m$ to $E$ over an on-chain exchange. 

We refer to a transaction that aims to copy and replace the actions of \txv as an \emph{imitation transaction} \txc. \txv interacts with a set of contracts \scvis, where \adversary may need to replace a subset of those contracts to achieve the successful execution of \txc. We denote that \adversary may process any victim contract \scvi to synthesize respective adversarial contract \scai.

\emph{Clarification.} \attackname cannot synthesize new attacks because \attackname has no prior knowledge of previous blockchain attacks and no knowledge of the application level logic. Therefore, \attackname is reliant on a template transaction and the associated contracts encoding attack logic.

\begin{figure*}[t]
    \centering
    \includegraphics[width=\textwidth]{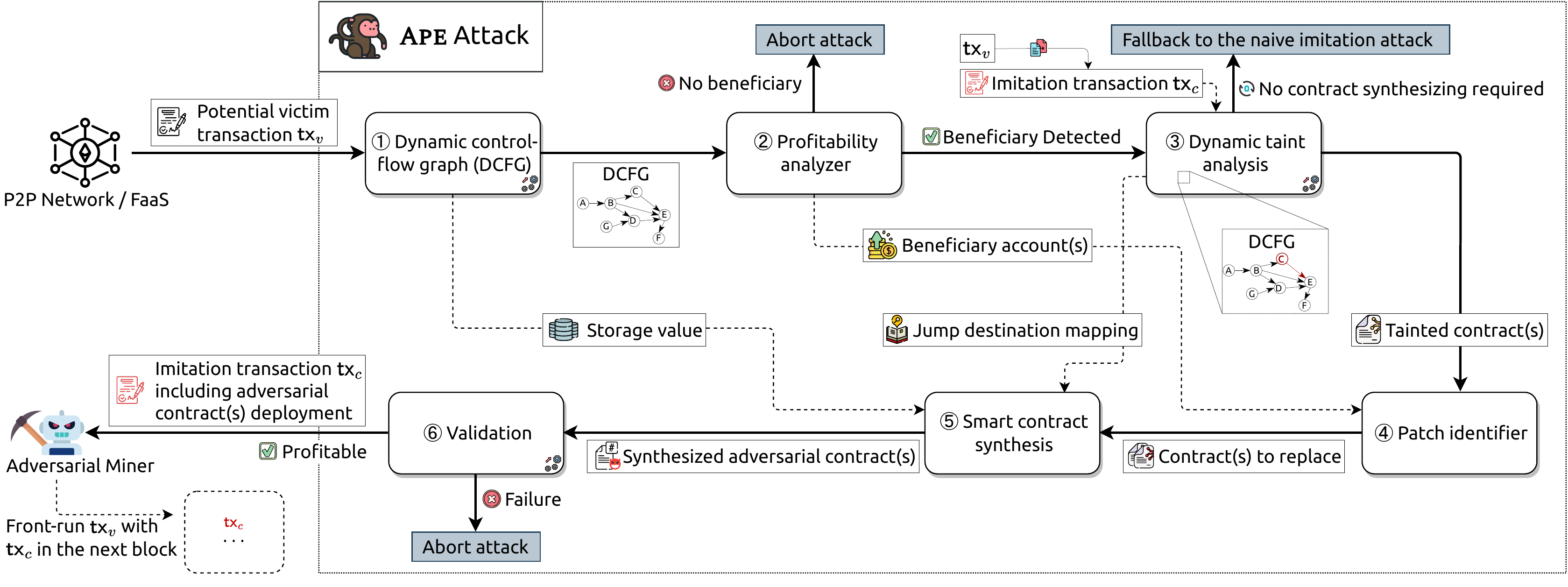}
    \caption{Overview of the \attackname, a real-time imitation attack on \EVM based blockchain transactions.}
    \label{fig:overview}
\end{figure*}

\subsection{Attack Overview}
Given a profitable victim transaction \txv, an adversary \adversary can attempt to imitate its logic with a naive imitation transaction (cf.\ Section~\ref{sec:transaction-replay-attack}). However, the naive method may fail due to various existing defense mechanisms, e.g., an authentication (cf.\ Section~\ref{sec:motivating-example}). Understanding a failure is challenging for \adversary due to the lack of prior knowledge about \txv. Moreover, the execution of \txv may be intertwined with multiple invoked contracts \scvis, complicating the analysis. Finally, \adversary must attack in real-time because the attack window typically only lasts a few seconds until \txv is mined. 

The objective of \attackname is to generate an imitation transaction \txc, as well as a set of synthesized contracts \scais that imitate the logic of \txv (and the associated victim contracts \scvis) with the following properties:

\begin{description}
\item[No prior victim knowledge] Any profitable transaction should be considered a potential victim, independent of the issuer. \adversary is not expected to have prior knowledge of the victim transaction, intent, or past blockchain history.
\item[No reasoning about the victim logic] \adversary has no knowledge about the application logic of \txc and involved \scvi.
\item[Higher payoff over additional complexity] \attackname introduces additional complexity over the naive imitation method~\cite{qin2022quantifying} and should therefore exceed its revenue. For an objective comparison, we evaluate both imitation methods over the same blockchain transaction history.
\item[Real-time] We assume that \adversary only attacks pending victim transactions. Therefore, it is essential that \attackname is faster than the inter-block time (e.g., about~$13.3$ seconds on Ethereum, $3$ seconds on \BSC). Thus, it is necessary to prioritize execution speed over attack optimality~\cite{qin2021attacking}. Note that \adversary, as a miner, could choose to fork the blockchain over \txc, granting an extended attack time window, which is beyond the scope of this work.
\end{description}

The high-level logic of \attackname operates as follows. Given a potential victim transaction \txv, \adversary first attempts to imitate \txv by creating and executing \txc as a naive imitation transaction. If \txc's execution reverts, \adversary identifies the execution traces triggering the failure. \adversary then synthesizes \scai replacing the \scvi that led \txc to revert. To this end, \adversary copies the victim's contract(s) bytecode, and amends the instructions that prevent a successful imitation execution. Finally, \adversary locally validates \scai deployment(s) and \txc's profitability. Under the assumption that \adversary can front-run any competitor (cf.\ Section~\ref{sec:models}), \attackname is risk-free. Note that, leveraging the transaction composability, \adversary can operate \scai deployment(s) and imitation execution (i.e., \txc) atomically within a single transaction.

\attackname (cf.\ Figure~\ref{fig:overview}) consists of six key components:

\begin{description}
\item[Step \Circled{1}: Dynamic control-flow graph] \adversary executes \txv locally at the current blockchain state and builds a \DCFG. The \DCFG captures \textit{(i)} the inter-contract invocations of \scvi while \txv executes, and \textit{(ii)} the contract bytecode execution flow of invoked \scvi.
\item[Step \Circled{2}: Profitability analyzer] \adversary extracts the asset transfers triggered by \txv to observe the beneficiary account(s). \adversary then attempts to replace those accounts as in to become the beneficiary. If \adversary does not succeed in becoming the beneficiary, \adversary aborts the attack.
\item[Step \Circled{3}: Dynamic taint analysis] \adversary performs a dynamic taint analysis and compares the analysis outcome to the \DCFG constructed in Step~\Circled{1}.
If the comparison shows no difference, \adversary proceeds with the naive transaction imitation (cf.\ Section~\ref{sec:transaction-replay-attack}). Otherwise, given the dynamic taint analysis and comparison to the \DCFG, \adversary identifies the tainted basic blocks which prevent a successful execution of the naive imitation.
\item[Step \Circled{4}: Patch identifier] Given the detected beneficiary accounts and tainted smart contracts, \adversary proceeds to identify all smart contracts that need to be replaced.
\item[Step \Circled{5}: Smart contract synthesis] \adversary synthesizes \scai by copying bytecode from \scvi. \adversary may need to amend the bytecode of \scai to ensure that \txc can execute \txv's logic and collect the produced financial revenue.
\item[Step \Circled{6}: Validation] \adversary deploys \scai locally and executes \txc to validate if the attack is profitable. If profitable, \adversary deploys \scai on-chain and issues \txc, front-running \txv.
\end{description}

\section{\attackname Details}
In this section, we present the design details of \attackname and discuss the technical limitations.
\subsection{Step \Circled{1}: Dynamic Control-Flow Graph}\label{sec:dynamic-control-flow-graph}
A smart contract \CFG~\cite{allen1970control} is a graph representation of the contract bytecode. In a \CFG, each node denotes a basic block, a linear sequence of instructions. Nodes are connected by directed edges, representing the code jumps in the control flow. For \EVM bytecode, two opcodes, \texttt{JUMP} and \JUMPI, control the execution path of code blocks. \texttt{JUMP} is an unconditional jump to the destination taken from the stack, while \JUMPI is a conditional jump. A \CFG only includes static information about a contract, while a \DCFG is a specialized \CFG with dynamic information taken from a given execution.

\emph{Dynamic Control-Flow Graph Construction in \attackname.} \adversary executes \txv locally to build a \DCFG, representing the execution details of \txv. To reason about the control flow of \txv, \adversary records the condition value of every \JUMPI, as necessary for the dynamic taint analysis (step \Circled{3}, cf.\ Section~\ref{sec:dynamic-taint-analysis}). By capturing the opcodes for contract calls (i.e., \texttt{CALL}, \texttt{DELEGATECALL}, \texttt{STATICCALL}, and \texttt{CALLCODE}), \adversary identifies invocations across contracts. The constructed \DCFG thus tracks the execution of all smart contracts invoked in \txv. The constructed \DCFG captures the concrete execution of \txv rather than the complete representation of \scvi. That is helpful in this work's context because the unexecuted basic blocks remain irrelevant to imitating the victim transaction. Therefore, the resulting \scai (cf.\ step \Circled{5}, Section~\ref{sec:smart-contract-synthesis}) will likely have fewer opcodes than \scvi and hence reduce the attack cost.

\subsection{Step \Circled{2}: Profitability Analyzer}\label{sec:profitability-analyzer}
The profitability analyzer aims to filter out victim transactions which are unlikely to be profitable. Intuitively, an \attackname attempt is profitable if the adversarial revenue (measured in $E$) is greater than the required transaction fees (cf.\ Definition~\ref{def:profitable-attack}).
\begin{definition}[Profitable Condition]\label{def:profitable-attack}
Given a blockchain state $S$, an \attackname attack is profitable for \adversary \textit{iff} $\operatorname{Bal}_\mathcal{A}^E(S') - \operatorname{Bal}_\mathcal{A}^E(S) > 0$, where $S'$ is the blockchain state after the attack transaction is applied.
\end{definition}

In the simplest scenario, \adversary can infer the profitability of imitating \txv by examining the balance change of the transaction sender. However, the victim may transfer the revenue of \txv to another smart contract under its control, instead of transferring to the sender account. Therefore, imitating \txv is profitable only if \adversary captures the beneficiary recipient. To determine the profitability, it is thus necessary for the \adversary to identify the profit of every account involved in the execution of \txv. To this end, \adversary can extract the asset transfers from the \DCFG constructed in step~\Circled{1}. This extraction is straightforward through analyzing the \EVM logs defined in asset implementation standards (e.g., ERC20). We proceed to define a beneficiary account in the execution of \txv (cf.\ Definition~\ref{def:beneficiary-account}).

\begin{definition}[Beneficiary Account]\label{def:beneficiary-account}
All asset values are denominated in $E$. An account $a$ is considered a beneficiary \textit{iff} the amount that $a$ receives is greater than the amount that $a$ pays out within the execution of \txv.
\end{definition}

We measure the profitability only in the native cryptocurrency $E$ to normalize financial value comparisons. This implies that \adversary must exchange all received assets to $E$ atomically after imitating \txv.


Imitating \txv may yield a profit \textit{iff} there exists a beneficiary account in the execution of \txv. Specifically, there are two cases in which \adversary does not abort the attack:
\begin{enumerate}
    \item If the sender is a beneficiary account, other accounts are irrelevant to the profitability analyzer.
    \item Otherwise, if the sender is not a beneficiary account, the collective profit of other beneficiary accounts, minus the potential loss of the sender account must remain positive.
\end{enumerate}

\adversary then exports the beneficiary account to the patch identifier (step~\Circled{4}) for further analysis (cf.\ Section~\ref{sec:patch-identifier}). Note that this methodology may introduce false positives because the profitability analyzer does not consider if a transaction is attackable. For example, a transaction withdrawing assets from a wallet contract to the transaction sender is classified as profitable because the sender is a beneficiary account. However, it is unavailing to imitate the withdrawal transaction. Such false positives will be purged in the validation phase (step~\Circled{6}).

\subsection{Step \Circled{3}: Dynamic Taint Analysis}\label{sec:dynamic-taint-analysis}
Dynamic taint analysis~\cite{newsome2005dynamic} is a program analysis method, which tracks information flow originating from taint sources (e.g., untrusted input) as a program executes. 
Dynamic taint analysis operates with a taint policy explicitly determining \textit{(i)} what instructions introduce new taint, \textit{(ii)} how taint propagates, and \textit{(iii)} how tainted values are checked~\cite{schwartz2010all}. 

\emph{Dynamic Taint Analysis in \attackname.} \adversary proceeds to execute an imitation transaction \txc copied from \txv. The execution of \txc may fail, if \txc contains inconsistencies (e.g., a different transaction sender) when compared to \txv's execution (cf.\ Section~\ref{sec:motivating-example}). Therefore, when executing \txc, \adversary applies dynamic taint analysis to track where and how \txc's execution fails. \adversary considers opcodes, which may trigger inconsistent execution values, as taint sources and tracks their taint propagation. We outline the taint analysis policy of \attackname in the following.


\begin{table}[t]
\small
\centering
\caption{Taint introduction rules. \texttt{ORIGIN} certainly introduces an inconsistent value when executing \txc, while the remaining opcodes (\crule{0.25cm}{0.25cm}) might, or might not, impact \txc's execution.}
\label{tab:taint-seed}
\begin{tabular}{ll}
\toprule
Taint Source & Description \\ \midrule
    \texttt{ORIGIN}                     & \txc sender address  \\
    \rowcolor{Gray}\texttt{CALLER}                     & Message caller address           \\
    \rowcolor{Gray}\texttt{ADDRESS} & Address of the executing contract     \\ 
    \rowcolor{Gray}\texttt{CODESIZE} & Length of the executing contract's code     \\
    \rowcolor{Gray}\texttt{SELFBALANCE} & Balance of the executing contract     \\
    \rowcolor{Gray}\texttt{PC} & Program counter     \\
    \bottomrule
\end{tabular}
\end{table}

\begin{description}
\item[Taint Introduction] We inspect all \EVM opcodes and identify those which may introduce inconsistencies (cf.\ Table~\ref{tab:taint-seed}). For example, \texttt{ORIGIN} (the transaction sender address) certainly produces an inconsistent value because \txc is issued from the adversarial, instead of the victim address. The remaining opcodes might, or might not, introduce inconsistencies during execution.
\item[Taint Propagation] When executing \txc, the taint propagates to the output of an arithmetic/logical operation, if at least one input is tainted. Notably, a storage variable is tainted if it is read from a tainted slot.\footnote{In the literature, this is referred to as a tainted address~\cite{newsome2005dynamic}. We, however, use the term ``tainted slot'' in this work to distinguish from ``contract address''.} 
\item[Taint Checking] Recall that we record the concrete value of every \JUMPI condition when building the \DCFG of \txv in step~\Circled{1} (cf.\ Section~\ref{sec:dynamic-control-flow-graph}). Given the tainted execution trace of \txc, we compare if every tainted \JUMPI is identical to the value recorded for \txv. This concrete comparison allows identifying how inconsistencies between \txv and \txc interrupt the execution of \txc.
\end{description}
We proceed to define a tainted basic block (cf.\ Definition~\ref{def:tainted-basic-block}) and tainted contract (cf.\ Definition~\ref{def:tainted-contract}).

\begin{definition}[Tainted Basic Block]\label{def:tainted-basic-block}
A basic block is tainted \emph{iff} \textit{(i)} the basic block contains a $\mathsf{JUMPI}$ opcode with a tainted condition value, and \textit{(ii)} the condition values are different in the executions of \txc and $\mathsf{tx}_v$.
\end{definition}

\begin{definition}[Tainted Contract]\label{def:tainted-contract}
A smart contract $\hat{\mathsf{sc}_{v_i}}$ is tainted \textit{iff} the contract contains at least one tainted basic block.
\end{definition}

\begin{figure}[tb]
\centering


\lstset{numbers=none,language=Opcode,framexleftmargin=5mm}
\begin{subfigure}[h]{\columnwidth}
\lstinputlisting[lastline=3,frame=lrt,belowskip=0pt]{dis}
\lstinputlisting[backgroundcolor=\color{lightgray},firstline=4,firstnumber=4,lastline=8,frame=lr,aboveskip=0pt,belowskip=0pt]{dis}
\lstinputlisting[firstline=9,firstnumber=9,frame=lrb,aboveskip=0pt,belowskip=0pt]{dis}
\caption{Caller authentication bytecode snippet.}
\vspace{7pt}
\end{subfigure}
\begin{subfigure}[h]{\columnwidth}
\includegraphics[width=\columnwidth]{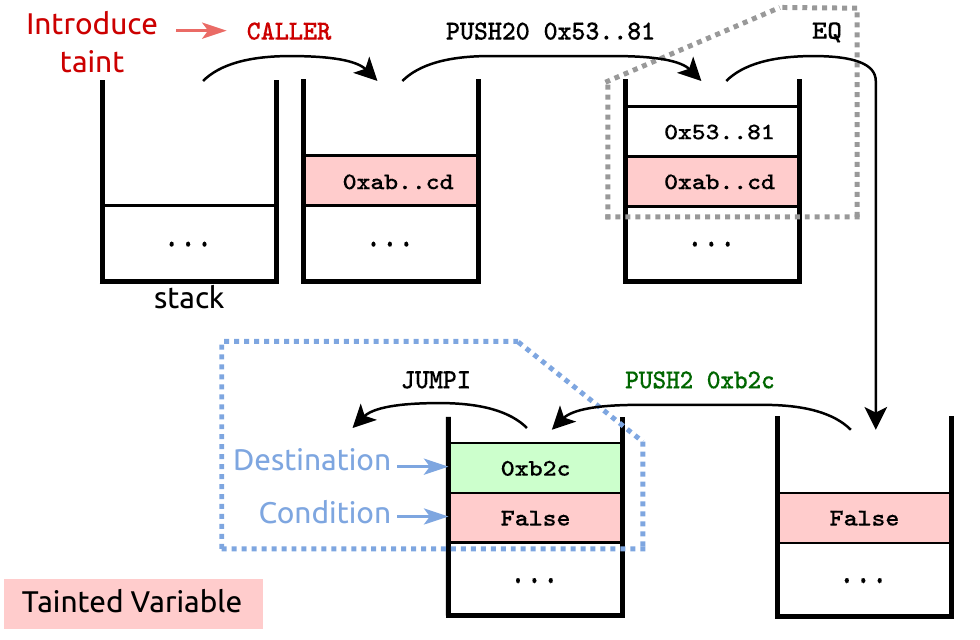}
\caption{Taint propagation visualization.}
\end{subfigure}
\caption{Taint propagation of the imitation transaction in the motivating example (cf.\ Figure~\ref{fig:non-replayable-transaction}, Section~\ref{sec:motivating-example}). \texttt{CALLER} introduces a tainted value, which propagates to a \JUMPI condition.}
\label{fig:taint_analysis}
\end{figure}

If there is no tainted basic block, the execution of \txv and \txc remain identical. Therefore, \attackname is then equivalent to the naive imitation attack (i.e., \adversary can imitate \txv by only issuing \txc and omits step \Circled{4}, \Circled{5}). However, if there exists a tainted basic block, the execution of \txc differs from \txv. To copy the execution of \txv, \adversary replaces tainted contracts with adversarial contracts, retaining the identical execution logic to \txv. 

\emph{Taint Propagation Example.} In Figure~\ref{fig:taint_analysis}, we showcase how the dynamic taint analysis tracks the execution of \txc in the motivating example (cf.\ Figure~\ref{fig:non-replayable-transaction}, Section~\ref{sec:motivating-example}), given an adversarial address \texttt{0xab..cd}. Following the execution, the taint propagates from \texttt{CALLER} to the condition value of \JUMPI (cf.\ PC \texttt{0xb27}, Figure~\ref{fig:taint_analysis}). Furthermore, the condition value is \texttt{False}, which is different from the execution of $\mathsf{tx}_v$. Therefore, $\mathcal{A}$ understands that the customized liquidation contract is tainted, and that it needs to be replaced.

\subsection{Step \Circled{4}: Patch Identifier}\label{sec:patch-identifier}
\begin{figure}[tb]
    \centering
    \includegraphics[width=\columnwidth]{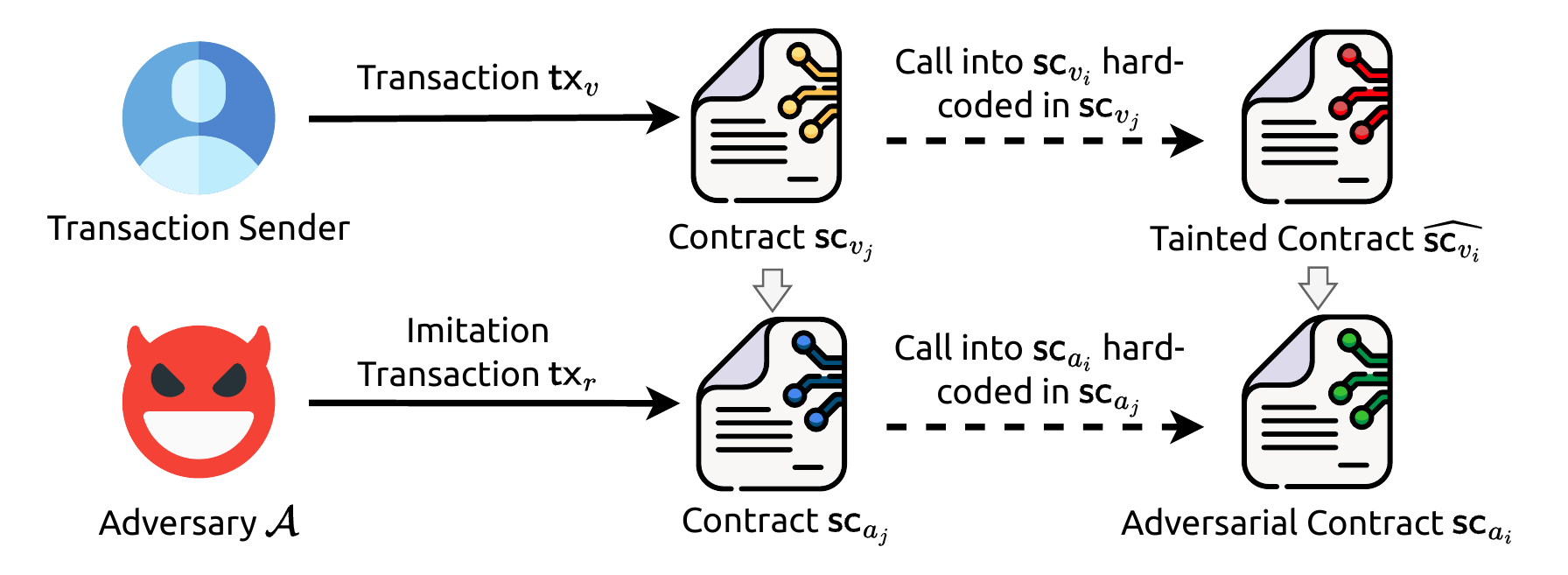}
    \caption{The invocation to the tainted contract \tscvi is hard-coded in the contract $\mathsf{sc}_{v_j}$. To perform \attackname, \adversary needs to replace both \tscvi and $\mathsf{sc}_{v_j}$.}
    \label{fig:patch-identifier}
\end{figure}

Recall that tainted basic blocks avoid \txc to successfully execute. Hence, \adversary attempts to replace the tainted contracts \tscvi with \scai. \adversary moreover needs to replace the beneficiary account identified in step~\Circled{2} so as to collect the financial revenue generated in \txc.
The patch identifier aims to detect all contracts out of \scvis that need to be patched and replaced to successfully imitate \txv. Depending on whether an invocation to \tscvi occurs from a transaction or from a contract, \adversary needs to distinguish the following two cases. We use \rscvi to denote a contract that must be patched and replaced.
\begin{description}
\item[Invocation from a transaction] When \txv is invoking \tscvi, \adversary should modify the \textit{to} address of \txc from \tscvi to \scai.
\item[Invocation from a contract] If the invocation to a tainted contract \tscvi is hard-coded (bytecode or storage) in a caller contract \scvj (cf.\ Figure~\ref{fig:patch-identifier}), \adversary should replace both \tscvi and \scvj to patch the hard-coded statement. This is necessary, even if \scvj is not tainted. The above patching procedure may apply iteratively to subsequently hard-coded contract invocations.
\end{description}

\subsection{Step \Circled{5}: Smart Contract Synthesis}\label{sec:smart-contract-synthesis}
\adversary proceeds to synthesize adversarial contract(s) to replace every \rscvi detected by the patch identifier (cf.\ Section~\ref{sec:patch-identifier}). 
On a high level, to synthesize \scai, \adversary copies bytecode from \rscvi with the following amendments.

For a tainted contract \tscvi, \adversary amends every tainted basic block to ensure that \scai follows the same code path as \tscvi, despite possible inconsistent \JUMPI conditions. Specifically, \adversary \textit{(i)} replaces \JUMPI with \JUMP, leading to an unconditional jump, or, \textit{(ii)} removes \JUMPI, leading to no jump. If the invocation to \tscvi is hard-coded in \scvj (cf.\ Figure~\ref{fig:patch-identifier}), \adversary needs to modify \scaj to redirect the contract invocation from \scaj to \scai. In addition, because every newly deployed adversarial contract has an empty storage, \scai may load an inconsistent value from its storage while \txc executes. \adversary hence further modifies \scai to recover the storage loading.

The aforementioned amendments ensure that \txc has the same execution path (code blocks and contracts) as \txv, but do not guarantee that \adversary receives the generated revenue. For example, the revenue generated in \txv may be sent to an account $a_v$, which is hard-coded in \scvi. The synthesized \scai copies the bytecode from \scvi and may follow the same asset transfer (i.e., to $a_v$ instead of an account controlled by \adversary). Therefore, to capture the generated revenue, \adversary needs to redirect the relevant asset transfers through modifying \EVM memory on the fly and injecting a revenue collection logic to \scai. Eventually, given the patched bytecode, \adversary updates jump destinations following the code size changes.

\subsection{Step \Circled{6}: Validation}\label{sec:validation}
Finally, \attackname locally validates \txc prior to the transaction being mined. \attackname could fail for two reasons: either \emph{(i)} the execution may fail, or \emph{(ii)} the financial revenue cannot cover the cost of deploying adversarial contracts and executing \txc.

To perform a concrete validation of the attack, a mining adversary deploys every \scai and executes \txc on the latest blockchain state locally. \adversary converts all received tokens to $E$ to check if \txc yields a profit. \txc can only yield a profit, if the revenue in $E$ covers all transaction fees including the smart contract(s) deployment fees. Recall that the adversarial contract(s) deployment, imitation execution, and asset exchange can be completed within one attack transaction. If the validation succeeds, \adversary includes the attack transaction, which front-runs \txv, in the next block. Otherwise, the attack is aborted, and \adversary bears no expenditure, i.e., \attackname is risk-free.

\begin{lstlisting}[float,floatplacement=H,label=lst:withdraw-with-sig,language=solidity, caption={Any account, providing an ECDSA signature signed by its private key, is allowed to withdraw assets from \texttt{Vault}.}]
contract Vault {
  function withdraw(bytes32 hash, uint8 v,
    bytes32 r, bytes32 s
  ) external {
    address signer = ecrecover(hash, v, r, s);
    if (msg.sender == signer) {
      msg.sender.transfer(
        address(this).balance);
    }
  }
}
\end{lstlisting}

\subsection{Limitations}
\attackname's design and implementation entails a number of limitations. For instance, we assume that the adversary has a sufficient amount of upfront assets required to execute \attackname successfully. Given the widespread access to flash loans, upfront capital requirements are solved~\cite{qin2021attacking}.

\attackname is not applicable when sophisticated semantic reasoning is necessary. We present an illustrative example in Listing~\ref{lst:withdraw-with-sig}, where the contract \texttt{Vault} allows the withdrawal of assets by any account that provides a valid ECDSA signature $(\texttt{v}, \texttt{r}, \texttt{s})$ signed by its private key. The design of \texttt{Vault} renders a \texttt{withdraw} transaction vulnerable to imitation attacks because of this anyone-can-withdraw logic. Nonetheless, for the adversary to execute the attack automatically, it would require the automation of semantic comprehension and signature generation, which \attackname does not support.



Moreover, \attackname cannot imitate non-atomic strategies, i.e., spanning over multiple independent blockchain transactions. For example, given an asset exchange victim transaction, a sandwich attacker~\cite{zhou2021high} may create two adversarial transactions, extracting profit from the victim. The goal of \attackname is not to generate such an attack. However, if a sandwich adversary creates an atomic sandwich transaction wrapping the victim exchange, \attackname can successfully challenge this transaction.

In our work, we assume that the victim is not aware of an \attackname adversary, and the \attackname attack strategy in particular. If a victim is aware of \attackname, then the victim could redesign its smart contract to harden its transactions against an \attackname adversary. The attack approach presented in this paper works well in practice, as we show in Section~\ref{sec:evaluation}. We outline possible counter-attack strategies in Section~\ref{sec:countermeasures}.

\section{\attackname Historical Evaluation}\label{sec:evaluation}

\label{sec:historical-evaluation}
\emph{Implementation.} We implement \attackname in~$\numprint{6582}$ lines of Golang code. Further details can be found in Appendix~\ref{app:implementation-details}.

We proceed to evaluate how the \attackname imitation attack could have performed over a timeframe of \TimeDuration (from the~\StartDate to the~\EndDate)\footnote{Ethereum block~\ETHStartBlock to~\ETHEndBlock and \BSC block~\BSCStartBlock to~\BSCEndBlock.} on Ethereum and \BSC, the top two smart contract-enabled blockchains by market capitalization at the time of writing. We evaluate in total \potentialVictimTransactions and \BSCpotentialVictimTransactions past transactions on Ethereum and \BSC respectively.

\label{sec:historical-setup}

\subsection{Methodology and Setup} We consider every past transaction as a potential victim transaction on which we apply the \attackname pipeline. If the attack succeeds, we save the associated synthesized smart contract(s), along with the yielded revenue and execution costs (e.g., gas cost for contract deployment and imitation transaction). If no contract replacement is required, \attackname falls back to the naive imitation attack, which we present separately as a baseline.

While a full archive node can provide the blockchain state at any past block, it does not directly allow the execution of arbitrary transactions on an arbitrary past blockchain state. We therefore implement an \emph{emulator} for both Ethereum and \BSC by customizing \href{https://github.com/ethereum/go-ethereum}{go-ethereum} and \href{https://github.com/bnb-chain/bsc}{bsc} \EVM accordingly. The emulator fetches historical states from an archive node and returns the execution result of any given transaction.
We perform the experiments on Ubuntu~$20.04.3$ LTS, with an AMD~$3990$X ($64$ cores),~$256$GB RAM and~$8$TB NVMe SSD.

\subsection{Evaluation Results}\label{sec:overall-statistics}
On Ethereum, from a total of~\potentialVictimTransactions potential victim transactions mined on-chain, we identify~\naiveReplayableTransactions (\naiveReplayablePercentage) vulnerable to the naive imitation attack (cf.\ Table~\ref{tab:attack-statistics}). \attackname successfully attacks~\newAttackTransactions (\newAttackTransactionPercentage) victim transactions, which involves replacing~\newAttackContracts unique smart contracts.

From the~\BSCpotentialVictimTransactions \BSC transactions, we discover that the naive imitation is applicable to~\BSCnaiveReplayableTransactions (\BSCnaiveReplayablePercentage) victim transactions, while \attackname captures~\BSCnewAttackTransactions (\BSCnewAttackTransactionPercentage) additional transactions, involving~\BSCnewAttackContracts unique contracts.


\begin{table}[h]
\centering
\caption{Overall attack statistics representing the successfully attacked transactions and unique victim contracts.}
\label{tab:attack-statistics}
\resizebox{\columnwidth}{!}{%
\begin{tabular}{cccccc}
\toprule
Chain & Attack       & Transactions & Contracts & Overall Profit (USD) & Average Profit (USD) \\ \midrule
\multirow{2}{*}{Ethereum} & Naive          &    \naiveReplayableTransactions     &   NA      &  \naiveReplayProfitNoUnit & \naiveReplayProfitAverageNoUnit \\
& \attackname &      \newAttackTransactions     &    \newAttackContracts    &  \newAttackAdditionalProfitNoUnit & \newAttackProfitAverageNoUnit   \\\midrule
\multirow{2}{*}{\BSC} & Naive  & \BSCnaiveReplayableTransactions     &   NA      &  \BSCnaiveReplayProfitNoUnit & \BSCnaiveReplayProfitAverageNoUnit \\
& \attackname &      \BSCnewAttackTransactions     &    \BSCnewAttackContracts    &  \BSCnewAttackAdditionalProfitNoUnit & \BSCnewAttackProfitAverageNoUnit   \\ \bottomrule
\end{tabular}%
}
\end{table}



\begin{figure}[t]
\centering
\includegraphics[width=\columnwidth]{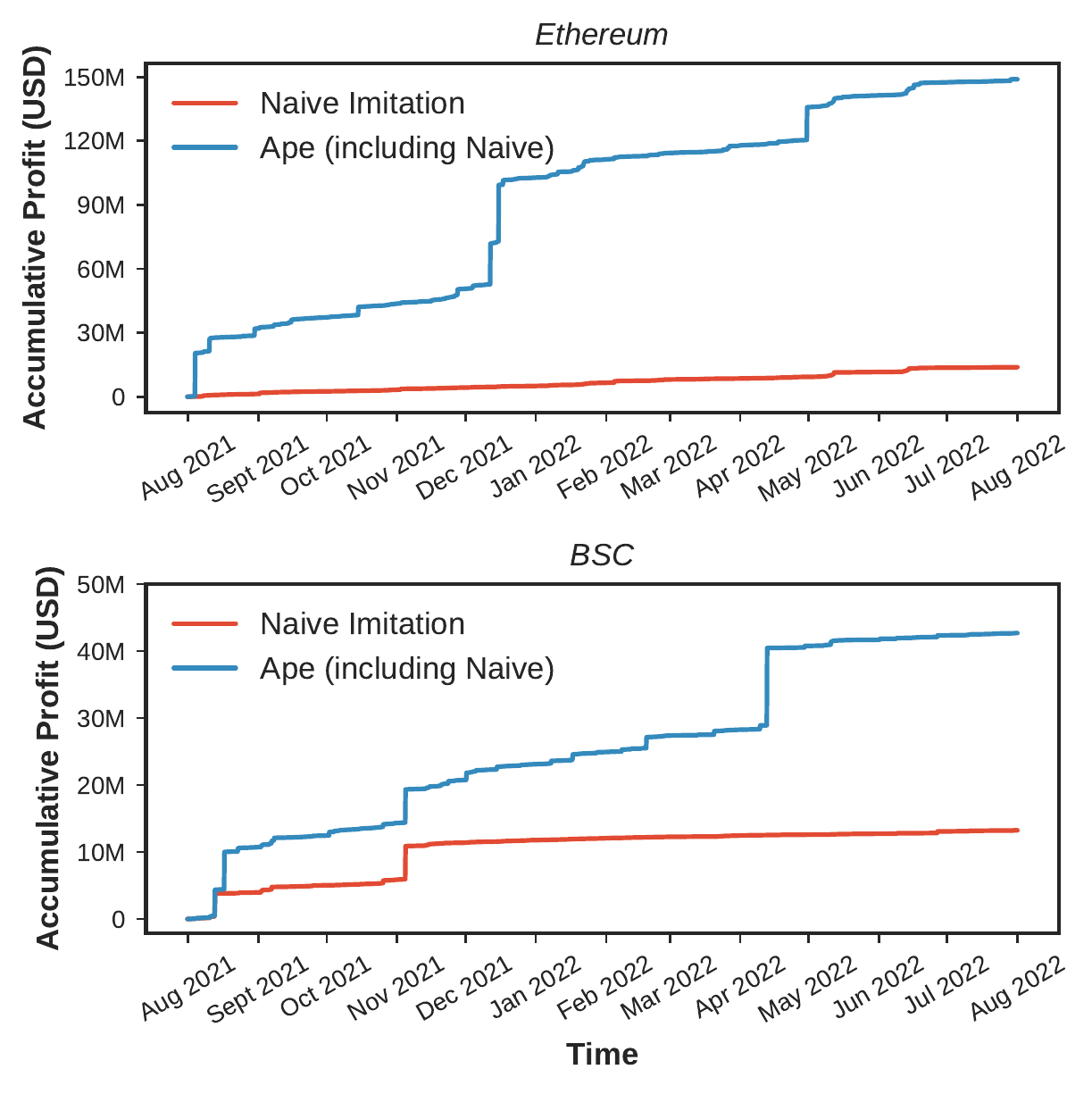}
\caption{From the~\StartDate to the~\EndDate, the total imitation attack profit (including \attackname and the naive imitation) on Ethereum reaches~\newAttackProfit, while the accumulative profit of \attackname is~\newAttackAdditionalProfit. On \BSC, \attackname and the naive imitation generate~\BSCnewAttackAdditionalProfit and~\BSCnaiveReplayProfit respectively.
}
\label{fig:accumulative-profit}
\end{figure}

\begin{figure}[t]
    \centering
    \includegraphics[width=\columnwidth]{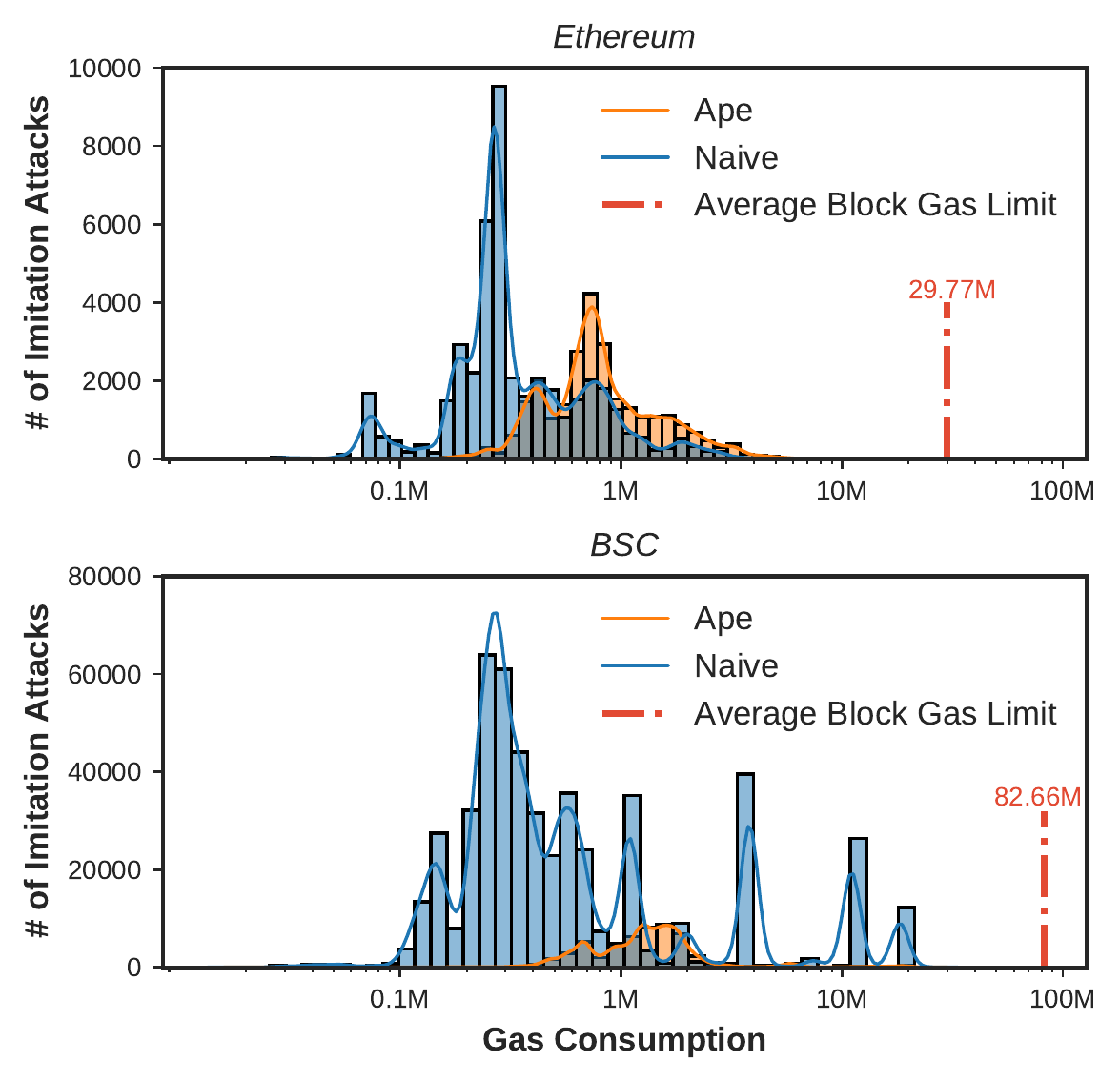}
    \caption{On average, \attackname on Ethereum costs~\newAttackGasConsumption gas per attack, while a naive imitation attack costs~\naiveReplayGasConsumption gas. The naive imitation attack on \BSC, however, has a higher average gas consumption (\BSCnaiveReplayGasConsumption) than the \attackname attack (\BSCnewAttackGasConsumption).}
    \label{fig:gas-consumption}
\end{figure}

\paragraph{Attack Profit} To reasonably measure the attack profit, we need to consider the transaction fee cost required by the contract deployment and imitation transaction. Because block space is limited (e.g., by the Ethereum block gas limit), when performing an \attackname attack, we need to capture the opportunity cost of a miner's forgoing transaction fees when not including a victim transaction. We therefore quantify such an opportunity cost with the transaction fee of \attackname's victim transaction.
The profit is converted to USD with the ETH (BNB) price at the time of each victim transaction. Note that ETH (BNB) is the native cryptocurrency on Ethereum (\BSC).

We present the accumulative profit of \attackname in Figure~\ref{fig:accumulative-profit}. On Ethereum, when compared to the naive imitation, \attackname could have increased the imitation attack profit by~\profitIncreaseRatio. Specifically, we find that \attackname (including the naive imitation attack) generates in total a profit of~\newAttackProfit, while the accumulative profit of the naive imitation attack over the same timeframe only amounts to~\naiveReplayProfit. Quantifying the required upfront capital for the imitation attacks, we find that~\UpfrontLessThanFiveEtherPercentage of the successful attacks require less than $5$ ETH (including transaction fees).

On \BSC, the naive imitation attack profit accumulates to~\BSCnaiveReplayProfit, while \attackname contributes an additional profit of~\BSCnewAttackAdditionalProfit (+\BSCprofitIncreaseRatio). \UpfrontLessThanFiveBNBPercentage of the \BSC imitation attacks require an upfront capital less than $5$~BNB.

We notice that \attackname captures an order of magnitude higher average profit compared to the naive imitation (cf.\ Table~\ref{tab:attack-statistics}). To further understand the transactions vulnerable to \attackname, we provide an analysis in Section~\ref{sec:analysis}.

\paragraph{Gas Consumption}
\attackname introduces additional gas costs because the adversary may need to deploy adversarial contracts. We identify two gas-related constraints by which an attack is bound. The revenue from the attack must cover the gas expenditures, and, the gas used to deploy and execute the attack transactions should remain below the block gas limit.



On Ethereum, we find that \attackname costs~\newAttackGasConsumption gas on average, while the naive imitation costs~\naiveReplayGasConsumption gas (cf.\ Figure~\ref{fig:gas-consumption}). As a reference, at the time of writing, Ethereum applies a dynamic block gas limit with an average of~\blockLimitAverage. The maximal gas consumption of \attackname on the identified historical transactions is~\newAttackGasConsumptionMax, which is below the average Ethereum block gas limit. On average, the adversarial contract deployment amounts to \newAttackGasContractDeploymentRatio of the attack gas consumption.

On average, \BSC has a higher block gas limit (\BSCblockLimitAverage) than Ethereum, which allows more space for imitation attacks. We observe that both \attackname (\BSCnewAttackGasConsumption) and the naive imitation (\BSCnaiveReplayGasConsumption) have a higher average gas consumption on \BSC. The adversarial contract deployment costs on average \BSCnewAttackGasContractDeploymentRatio of the \attackname gas consumption on \BSC. Contrary to Ethereum, the naive imitation on \BSC has a higher average gas consumption than \attackname. We therefore analyze the~\BSCGasOverThreeM naive imitation transactions which cost more than~$3$M of gas. We identify in total~\BSCGasToken transactions that are related to the gas token minting event.\footnote{\BSC \EVM allows consuming gas to mint so-called gas tokens (e.g., \bscscanTx{0x6bdcc83369ac7f04f898b57330d6f496b0f018d53be9f8fa9f92e02acfa1c07b}), which are tradable, hence creating arbitrage opportunities.} After removing the~\BSCGasToken outliers, the average gas cost of the naive imitation on \BSC is~\BSCnaiveReplayGasConsumptionWOGasToken.



\paragraph{Adversarial Contract}
On average, an \attackname attack requires replacing~\averageContractToReplace contracts on Ethereum and~\BSCaverageContractToReplace contracts on \BSC. We find that every synthesized adversarial contract is on average~\contractSizeReduction smaller (in bytes) compared to the replaced victim contract (cf.\ Table~\ref{tab:adversarial-contract-statistics}). Because \attackname may expand victim contracts with synthesized code, we also observe a negative reduction of~\contractSizeReductionMin, a worst-case increase of~\contractSizeReductionMinNegative. On \BSC, the average contract size reduction is~\BSCcontractSizeReduction, with a maximum increase of~\BSCcontractSizeReductionMinNegative.

\begin{table}[t]
\centering
\caption{Adversarial contract statistics. \attackname creates at most \averageContractToReplaceMax adversarial contracts on both Ethereum and \BSC, with an average of \averageContractToReplaceMean and \BSCaverageContractToReplaceMean respectively. 
}
\label{tab:adversarial-contract-statistics}
\resizebox{\columnwidth}{!}{%
\begin{tabular}{cccccc}
\toprule
                &            & Mean & Std. & Max & Min \\ \midrule
\multirow{2}{*}{Ethereum} & Adversarial Contract Number &  \averageContractToReplaceMean   &  \averageContractToReplaceStd   &  \averageContractToReplaceMax    & \averageContractToReplaceMin     \\
& Contract Size Reduction     & \contractSizeReductionMean    &  \contractSizeReductionStd   & \contractSizeReductionMax     &  \contractSizeReductionMin    \\ \midrule
\multirow{2}{*}{\BSC} & Adversarial Contract Number &  \BSCaverageContractToReplaceMean   &  \BSCaverageContractToReplaceStd   &  \BSCaverageContractToReplaceMax    & \BSCaverageContractToReplaceMin     \\
& Contract Size Reduction    & \BSCcontractSizeReductionMean    &  \BSCcontractSizeReductionStd   & \BSCcontractSizeReductionMax     &  \BSCcontractSizeReductionMin    \\ \bottomrule
\end{tabular}%
}
\end{table}

\subsection{Historical Analysis}\label{sec:analysis}
To distill insights from \attackname's success, we manually investigate the top-\AnalyzedVictims most rewarding \attackname victims on Ethereum and \BSC, capturing a profit of~\AnalyzedVictimTotalProfit (\AnalyzedVictimTotalProfitPercentage) and~\BSCAnalyzedVictimTotalProfit (\BSCAnalyzedVictimTotalProfitPercentage) respectively. The overall transaction and profit distributions are presented in Table~\ref{tab:analysis-distribution}. We proceed to outline the details of our findings.

\begin{table}[t]
\centering
\caption{Transaction and profit distributions of the top-\AnalyzedVictims most rewarding \attackname victims on Ethereum and \BSC. We fail to classify~$31$ \BSC victim transactions.}
\resizebox{\columnwidth}{!}{%
\begin{tabular}{@{}ccccccc@{}}
\toprule
\multirow{2}{*}{Category} & & \multicolumn{2}{c}{Ethereum} &  & \multicolumn{2}{c}{BSC} \\ \cmidrule(lr){3-4} \cmidrule(l){6-7} 
                         & & transactions          & Profit (USD)         &  & transactions       & Profit (USD)       \\ \cmidrule(r){1-1} \cmidrule(lr){3-4} \cmidrule(l){6-7} 
    Arbitrage \& Liquidation     &                 &    \ArbLiqTransactions         &  \ArbLiqProfitNoUSD              &  &      \BSCArbLiqTransactions    &    \BSCArbLiqProfitNoUSD          \\
    Known \DeFi Attacks    &                 &    \DeFiAttackTotalTransactions         &  \DeFiAttackProfitNoUSD              &  &      \BSCDeFiAttackTotalTransactions    &    \BSCDeFiAttackProfitNoUSD          \\
    Known Vulnerabilities    &                 &    \knownVulTransactions         &  \knownVulProfitNoUSD            &  &      --    &    --         \\
    Newly Found Vulnerabilities  &                 &  \newVulTransactions &  \newVulProfitNoUSD      &  &     \BSCNewVulTransactions    &   \BSCNewVulProfitNoUSD          \\
    \bottomrule
\end{tabular}%
}
\label{tab:analysis-distribution}
\end{table}





\paragraph{Known DeFi Attacks and Vulnerabilities}
On Ethereum, \attackname discovers~\DeFiAttackBlackhatTransactions blackhat transactions and \DeFiAttackWhitehatTransactions proclaimed whitehat transactions corresponding to~\DeFiAttacks known \DeFi attacks, capturing a total profit of~\DeFiAttackProfit. The most profitable \attackname vulnerable transaction \abbrEtherscanTx{0xcd7dae143a4c0223349c16237ce4cd7696b1638d116a72755231ede872ab70fc} is a DeFi attack on the Popsicle Finance smart contracts, generating an \attackname profit of \PopsicleProfit~USD. Note that this \attackname profit is lower than the reported attack profit, because we convert \attackname's revenue to ETH. An exchange that may incur excessive slippage, particularly for higher amounts.


We detect two transactions that might be related to two disclosed contract vulnerabilities. Transaction \abbrEtherscanTx{0x2e7d7e7a6eb157b98974c8687fbd848d0158d37edc1302ea08ee5ddb376befea} and \abbrEtherscanTx{0xe12ae015c8023bbe6405662a3ddf5e8e106e7f6255e905b7312dcf65b27d755c}, respectively involve the victim contracts reported in \href{https://blog.auctus.org/action-required-critical-vulnerability-3d448d4d0dcb}{Auctus ACOWriter vulnerability} and \href{https://twitter.com/peckshield/status/1509009746744983556}{BMIZapper vulnerability}. However, we could not find the contract operators disclosing public information about these transactions.

On \BSC, \attackname captures~\BSCDeFiAttacks \DeFi attacks involving~\BSCDeFiAttackTotalTransactions transactions, generating a total imitation profit of~\BSCDeFiAttackProfit.

Details of the identified DeFi attacks and vulnerabilities are outlined in Table~\ref{tab:attacks-and-vulnerabilities} and~\ref{tab:bsc-attacks-and-vulnerabilities}, Appendix~\ref{app:evaluation-analysis}. Note that for each attack transaction, we check whether \href{https://etherscan.io/}{etherscan.io} observes the attack on the \PtoP network. If etherscan would not detect the attack on the \PtoP network, the attack is likely being propagated privately to miners (e.g., \FaaS). We find that~\DeFiAttackWhitehatTransactionsPrivate of the~\DeFiAttackWhitehatTransactions whitehat attack transactions (corresponding to six attacks) are propagated to miners privately. Whitehat hackers may perfer private communication channels for two purposes: \textit{(i)}~to accelerate the whitehat transaction inclusion on-chain; \textit{(ii)}~to mitigate the possibility of imitation attacks. 

\paragraph{Newly Found Vulnerabilities}
\attackname uncovers five undisclosed vulnerabilities. Among these vulnerabilities, the most profitable one, named \emph{massDeposit}, generates a total profit of~\MassDepositLoss and~\BSCMassDepositLoss on Ethereum and \BSC respectively. A case study of the massDeposit vulnerability is presented in the following, while the details of other newly found vulnerabilities are provided in Appendix~\ref{app:evaluation-analysis}.

\emph{Responsible Disclosure.} At the time of writing, all vulnerabilities discovered have no more active funds to be exploited. Yet, new users may interact with such contracts unknowingly despite the present danger. Therefore, we choose to attempt to contact the vulnerable smart contracts through a dedicated blockchain messaging service (\href{https://chat.blockscan.com/start}{Blockscan Chat} by Etherscan). Unfortunately, all the vulnerable contracts we identified appear to be anonymously deployed and there is no apparent means to identify an entity or person behind these contracts.

\begin{lstlisting}[float,floatplacement=H,label=lst:depositer-contract,language=solidity, caption={\attackname vulnerable \href{https://etherscan.io/address/0xe2c071e1e1957a62fddf0199018e061ebfd3ac2c\#code}{\texttt{Depositer} contract}, where anyone can invoke the \texttt{massDeposit()} function and propose a vault contract. Our evaluation shows that \attackname could have caused a potential loss of \MassDepositLoss to this contract.}]
contract Depositer {
  address private owner;
  function massDeposit(
    VaultV0 vault, IERC20 token,
    address[] calldata lst,
    uint[] calldata amt
  ) external {
    token.approve(address(vault), 2 ** 256-1);
    require(lst.length == amt.length);
    for (uint i = 0; i < lst.length; i++) {
      vault.depositOnBehalf(lst[i], amt[i]);
    }
    vault.setOwner(owner);
  }
}
\end{lstlisting}

\begin{figure}[t]
\centering
\begin{subfigure}[t]{\columnwidth}
    \includegraphics[width=\columnwidth]{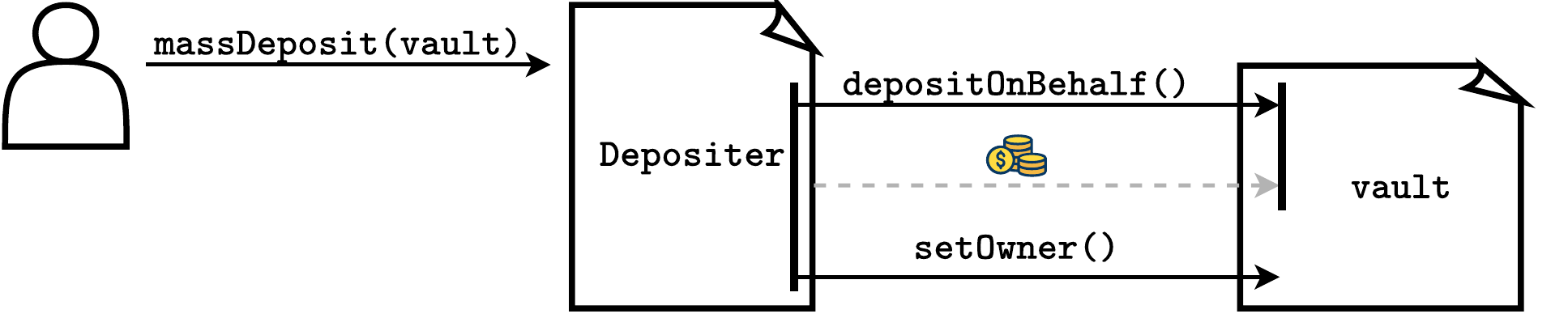}
\caption{\txv: \texttt{vault} collects assets from \texttt{Depositer}.}
\label{fig:massdeposit-victim}
\end{subfigure}%
\vfill
\vspace{5pt}
\begin{subfigure}[t]{\columnwidth}
\includegraphics[width=\columnwidth]{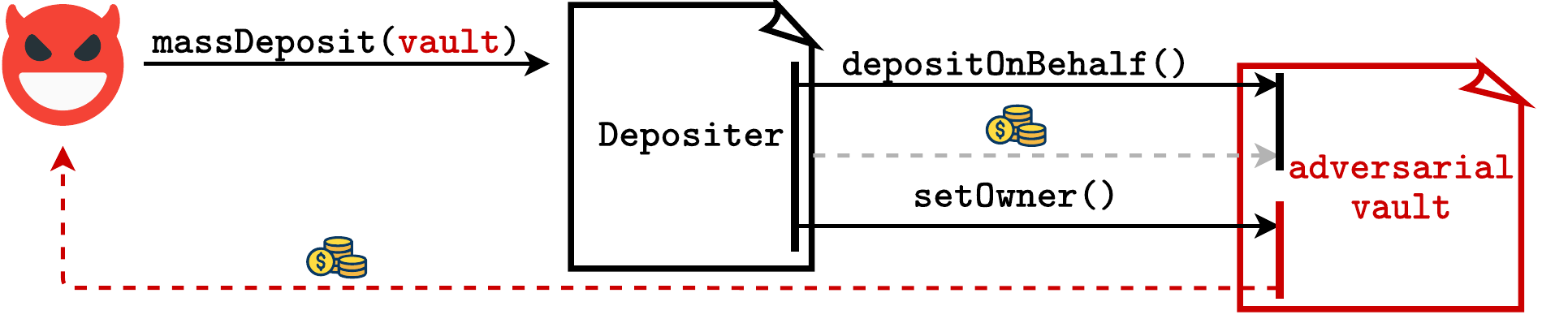}
\caption{\txc: \attackname automatically synthesizes an adversarial vault, which transfers collected assets to the adversarial account by the end of the transaction.}
\label{fig:massdeposit-autoplay}
\end{subfigure}
\caption{The massDeposit vulnerability.}
\label{fig:massdeposit}
\end{figure}

\emph{massDeposit Case Study.}
The massDeposit vulnerability exists with a \texttt{Depositer} contract (\etherscanAddress{0xe2c071e1E1957A62fDDf0199018e061ebFD3ac2C}) on Ethereum (cf.\ Listing~\ref{lst:depositer-contract}).\footnote{on \BSC, we also discover a contract (\bscscanAddress{0xA9F22770dbF9d19D49Bd63ea918eE8c9c77dB016}) with the identical vulnerable code, from which up to~\BSCMassDepositLoss could have been exploited during our evaluation period.} Within our evaluation period, the \texttt{Depositer} contract received deposits worth~\MassDepositLoss, later collected by a \texttt{vault} contract. We find that the deposited funds could have been stolen through an \attackname attack due to the following vulnerability details.
The \texttt{massDeposit()} function of the \texttt{Depositor} contract takes as argument a \texttt{vault}, which is the contract receiving the totality of the \texttt{Depositer} contract's funds (cf.\ Figure~\ref{fig:massdeposit-victim}). Anyone can call the function \texttt{massDeposit()}, and provide an arbitrary vault contract address. The \texttt{vault} contract is then allowed to collect assets from the \texttt{Depositer} contract. Eventually, \texttt{Depositer} invokes the \texttt{setOwner()} function of \texttt{vault}, attempting to configure the ownership of \texttt{vault}. When observing such a \texttt{massDeposit} transaction, an \attackname attacker \adversary identifies \texttt{vault} as the beneficiary account. \attackname then automatically synthesizes an adversarial contract to replace the \texttt{vault} contract (cf.\ Figure~\ref{fig:massdeposit-autoplay}). Recall that when synthesizing a beneficiary contract, \attackname injects the logic to transfer all collected assets to the adversarial account by the end of the imitation transaction.
Therefore, \adversary extracts the assets from \texttt{Depositer} after executing the imitation transaction. It should be noted that, even without \attackname, an attacker can manually craft an adversarial \texttt{vault} contract to maliciously extract assets from \texttt{Depositer}. We hence classify this contract design as a vulnerability.

\section{\attackname Real-time Evaluation}
Ideally, an \attackname-enabled miner attempts to identify in real-time the optimal transaction order for its mined blocks. As such, given a list of unconfirmed transactions, the miner could apply every combination of transactions to identify \attackname's optimal revenue. However, exploring such combinations will soon result in a combinatorial explosion.

Therefore, we shall come up with a more realistic, yet best-effort solution for the miner's transaction ordering. We can borrow the widely adopted assumption~\cite{zhou2021high}, that transactions are typically sorted by their fee amount within a blockchain block. As such, in a fee-ordered list, we know for a victim transaction $v_i$, that the \attackname transaction should also be placed at position $i$. While the \EVM currently only supports the sequential execution of transactions, promising efforts on speculative parallel execution~\cite{chen2021forerunner} could speed up the block validation as well as the victim transaction identification in \attackname. Such a parallel execution framework, however, is beyond the scope of this work.

\subsection{Methodology and Setup} To evaluate the real-time performance, we inject the \attackname logic into an Ethereum (\BSC) full node, referred to as an \attackname node. The \attackname node listens to the Ethereum (\BSC) \PtoP network and operates \attackname on every potential victim transaction without publishing the generated attack transaction. The repetitive evaluation process takes the current blockchain height $h$ and the memory pool $\mathcal{P}$ as the input (cf.\ Algorithm~\ref{alg:real-time-evaluation}). We first rule out the illegitimate transactions (e.g., a transaction with a wrong nonce number) from $\mathcal{P}$ and sort the resulting transactions in a descending gas price order. We then sequentially apply the sorted transactions on the current blockchain state. Given each pair of intermediate state $S_i$ (after applying $\mathsf{tx}_i$) and following transaction $\mathsf{tx}_{i+1}$, the \attackname node executes the attack function $\mathsf{Ape}(S_i, \mathsf{tx}_{i+1})$ asynchronously and continues with the next pair $(S_{i+1}, \mathsf{tx}_{i+2})$ in a non-blocking manner. If a transaction \txv is attackable, we replace (i.e., ``front-run'') \txv with the generated attack transaction. This pipeline corresponds to an adversarial miner \textit{(i)} selecting and sorting transactions from its memory pool, \textit{(ii)} applying every transaction sequentially to construct the next block, and \textit{(iii)} operating the \attackname attack when applicable. The \attackname node repeats this process and records the performance of the following metrics.

\newcommand\mycommfont[1]{\small\ttfamily\textcolor{gray}{#1}}
\SetCommentSty{mycommfont}

\begin{algorithm}[t]
\small
\SetAlFnt{\small}
\SetAlCapFnt{\small}
\SetAlCapNameFnt{\small}
\DontPrintSemicolon
\SetAlgoLined
\SetKw{KwForIn}{in}
\SetKwProg{Fn}{Function}{:}{end}
\SetKwFunction{FApe}{Ape}
\SetKwFunction{FFilter}{Filter}
\SetKwFunction{FSort}{Sort}
\SetKwFunction{FMetrics}{RealtimePerformanceMetrics}
\SetKwProg{Alg}{Algorithm}{:}{end}
\SetKwFunction{ARealTimeApe}{RealTimeApe}
\KwIn{The current blockchain height $h$;\ the local memory pool $\mathcal{P}$.
}
\Alg{\ARealTimeApe{$h$, $\mathcal{P}$}}{
    $S :=$ blockchain state at height $h$\;
    \ForEach{$\mathsf{tx}_v$ \KwForIn \FSort{\FFilter{$\mathcal{P}$}}}{
        \FApe{$S$, $\mathsf{tx}_v$} \tcp*{\FApe executes asynchronously}
        $S := \mathsf{tx}_v(S)$\;
    }
}\;
\Fn{\FApe{$S$, \txv, $h+1$}}{
    $t_0 := \operatorname{now}()$\;
    Apply the \attackname pipeline (cf.\ Figure~\ref{fig:overview})\;
    \If{\txv is vulnerable on $S$}{
        $t_1 := \operatorname{now}()$\;
        store \txv, $t_0$, $t_1$, $h+1$
    }
}\;
\Fn{\FMetrics{$t_0$, $t_1$, $h+1$}}{
    $t_2 := \text{block}\ h+1\ \text{arrival time}$\;
    $\text{\bf single transaction performance} := t_1 - t_0$\;
    $\text{\bf mempool performance} := t_2 - t_1$\;
}
\caption{\attackname Real-time Evaluation.}
\label{alg:real-time-evaluation}
\end{algorithm}

\newtheorem{metric}{Metric}
\begin{metric}[Single transaction performance]\label{metric:single-tx}
The time it takes \attackname to generate an attack given one victim transaction and the corresponding blockchain state (cf.\ Algorithm~\ref{alg:real-time-evaluation}).
\end{metric}

\begin{metric}[Mempool performance]\label{metric:mempool}
Given the dynamic mempool of transactions in real-time, the time from an \attackname attack generation to the arrival of the next block, i.e., the target block that should include \txc (cf.\ Algorithm~\ref{alg:real-time-evaluation}).
\end{metric} 


We execute the real-time evaluation with the same hardware setup as Section~\ref{sec:historical-setup} and a $10$ Gbps Internet connection. To swiftly capture pending transactions, we increase the \attackname node's maximum network peers from the default $50$ to $500$ for Ethereum and \BSC. The \attackname node spawns~$150$ threads executing the \texttt{Ape} function asynchronously (cf.\ Algorithm~\ref{alg:real-time-evaluation}).

\paragraph{Weaker Threat Model}
Note that contrary to the threat model in Section~\ref{sec:threatmodel}, in this section, we are not a miner. We further do not have private peering agreements with \FaaS and may therefore not observe victim transactions that only \FaaS-enabled miners would receive. Compared to the previous threat model, our \attackname node's adversarial capability in observing victim transactions is hence weaker. Therefore, in our real-time evaluation, we choose to focus on the computational real-time performance of \attackname in practice, ignoring the potential financial profit that the \attackname node could have generated under a weaker threat model. We expect weaker threat models to earn an inferior profit and hence leave such analysis to future work.


\begin{table}[b]
\centering
\caption{Single transaction performance (Ethereum) of the naive and \attackname imitation attack. Step \Circled{1}, \Circled{3}, and \Circled{6} accounting for~\StepOneThreeSixPercentage of the execution time of \attackname on average. \attackname shows an equivalent single transaction performance on \BSC.}
\label{tab:attack-time-breakdown}
\resizebox{\columnwidth}{!}{%
\begin{tabular}{lllll}
\toprule
                         & Mean (s) & Std.\ (s) & Max (s) & Min (s)\\ \midrule
Step~\Circled{1} DCFG                     &  \steponetimemean   & \steponetimestd    &  \steponetimemax    &   \steponetimemin   \\
Step~\Circled{2} Profitability Analyzer   &  \steptwotimemean   & \steptwotimestd & \steptwotimemax     &  \steptwotimemin    \\
Step~\Circled{3} Dynamic Taint Analysis   & \stepthreetimemean    & \stepthreetimestd    & \stepthreetimemax     & \stepthreetimemin \\
Step~\Circled{4} Patch Identifier         &  \stepfourtimemean   &  \stepfourtimestd   &     \stepfourtimemax & \stepfourtimemin     \\
Step~\Circled{5} Smart Contract Synthesis &  \stepfivetimemean   & \stepfivetimestd    & \stepfivetimemax     &   \stepfivetimemin   \\
Step~\Circled{6} Validation               & \stepsixtimemean    & \stepsixtimestd & \stepsixtimemax &  \stepsixtimemin\bigstrut\\\hdashline
Overall time cost of \attackname & \timeConsumptionMean & \timeConsumptionStd & \timeConsumptionMax & \timeConsumptionMin\bigstrut[t]\\\midrule
Overall time cost Naive Imitation & \naiveTimeConsumptionMean & \naiveTimeConsumptionStd & \naiveTimeConsumptionMax & \naiveTimeConsumptionMin \\\bottomrule
\end{tabular}%
}
\end{table}

\subsection{Computational Real-time Performance}
Our real-time experiment for Ethereum captures~\ETHRealtimeEffectiveTime of data. During the experiment, the Ethereum \attackname node receives~\ETHIncomingTxRate unique transactions per second on average. We detect in total~\ETHRealtimeTxBoth imitation opportunities, including~\ETHRealtimeTxNewAttack transactions vulnerable to the \attackname attack. The \BSC real-time evaluation spans~\BSCRealtimeEffectiveTime. Our \BSC \attackname node receives an average of~\BSCIncomingTxRate unique transactions per second. We identify~\BSCRealtimeTxNewAttack naive and~\BSCRealtimeTxNaive \attackname opportunities on \BSC.


Table~\ref{tab:attack-time-breakdown} presents the single transaction performance (cf.\ Metric~\ref{metric:single-tx}) and the execution time breakdown. On average, it takes~\naiveTimeConsumption seconds to generate a naive imitation attack, while generating an \attackname attack requires~\timeConsumption seconds. The most time-consuming steps of \attackname are \DCFG (step \Circled{1}), dynamic taint analysis (step \Circled{3}), and validation (step \Circled{6}), accounting for \StepOneThreeSixPercentage of the overall execution time.

\begin{figure}
    \centering
    \includegraphics[width=\columnwidth]{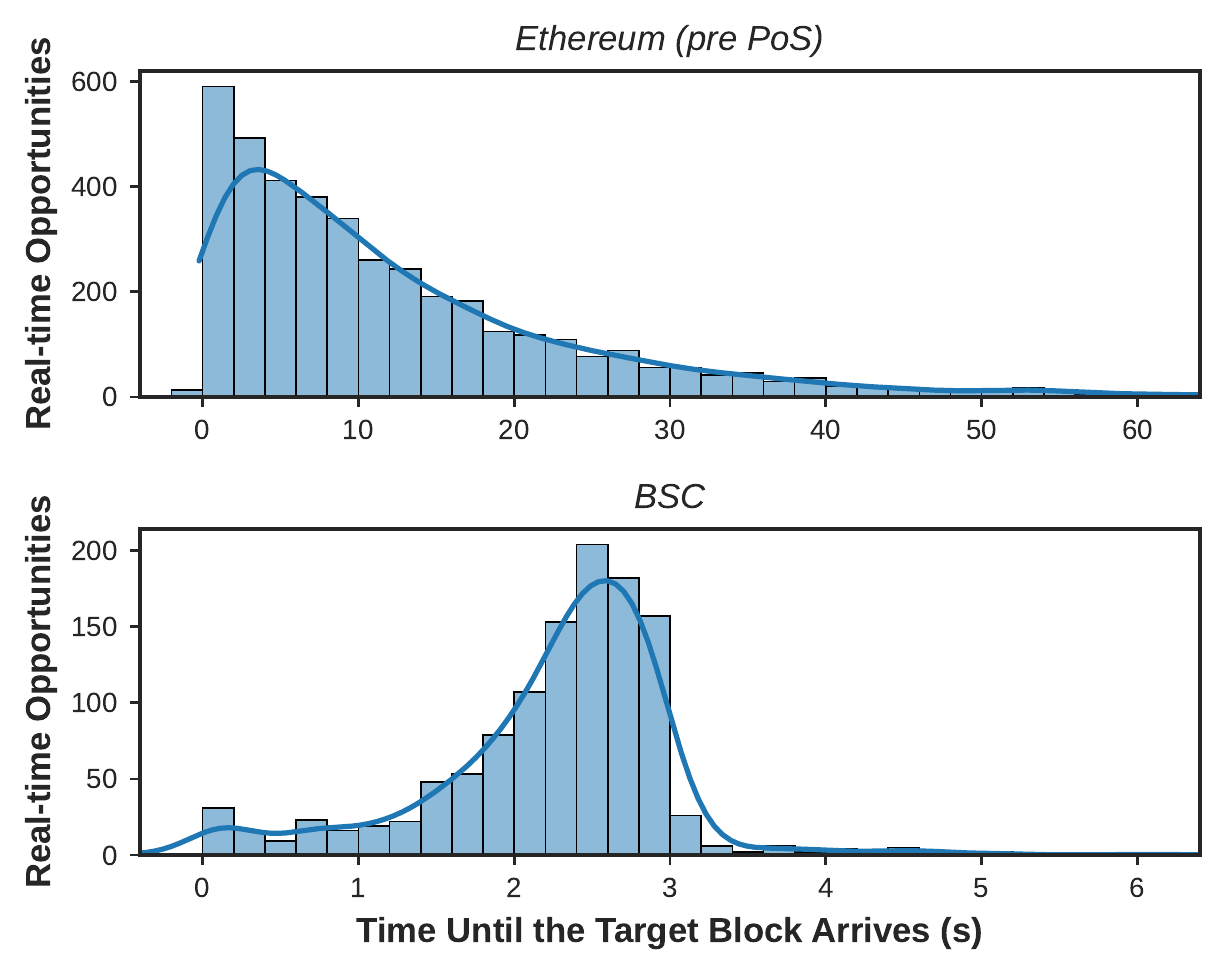}
    \caption{Distribution of the time duration from the attack generation to the target block reception (cf.\ Metric~\ref{metric:mempool}). \ETHRealtimeSuccessRate of the attacks on Ethereum and~\BSCRealtimeSuccessRate of the attacks on \BSC are generated before our \attackname node receives the next block. We notice that the time distribution on Ethereum resembles an exponential distribution~\cite{decker2013information} because at the time of the experiment Ethereum follows the Nakamoto consensus~\cite{bonneau2015sok,bano2019sok}, while \BSC has a less volatile block time interval.}
    \label{fig:opportunity_window}
\end{figure}

We show the mempool performance (cf.\ Metric~\ref{metric:mempool}) in Figure~\ref{fig:opportunity_window}. On Ethereum (BSC),~\ETHRealtimeSuccessRate (\BSCRealtimeSuccessRate) of the \attackname attacks are generated before our \attackname node receives the target block. The time duration from the attack generation to the target block reception is on average~\ETHRealtimeMempoolPerformance (\BSCRealtimeMempoolPerformance) seconds. Our evaluation results show the real-time property of \attackname on Ethereum, with a block interval of~$13.3$ seconds, and \BSC, with a block interval of~$3$ seconds.

\section{\attackname Countermeasures}\label{sec:countermeasures}
The imitation attacks pose an imminent threat to the \DeFi ecosystem and its users. \EVM-compatible blockchains rely on miners to execute and validate every transaction. It is therefore difficult to completely prevent a miner from inspecting and imitating a victim transaction. We consider the following countermeasures that possibly mitigate \attackname.


\paragraph{Imitation as a Defence Tool}
A possibly intuitive but effective alternative is to apply imitation as a defence mechanism. Miners could for example resort to automated whitehat hacking of \DeFi attacks by simply imitating attacks and refunding the proceeds to the victim. While miners are currently entrusted to secure the blockchain consensus, they can extend that capacity to the application layer. A bug bounty could incentivise the miners' altruism. We, however, leave the exploration of a precise incentive structure to future work.

\paragraph{Breaking Atomicity}
Breaking the atomicity of a profitable transaction can thwart \attackname, i.e., a user can split the logic of its transaction into multiple independent transactions. \attackname would then be incapable of capturing the entire logic from any separate transaction and would be unable to mimic it. Moreover, this defense can be implemented in a tool that would take a single transaction as input and automatically split it into multiple transactions with the same program logic. The primary shortcoming of such approach is that it does not apply to all cases (e.g., flash loan). Breaking the atomicity may also introduce additional risks to users (e.g., a partial arbitrage execution leads to a financial loss).

\paragraph{Front-running Mitigation}
As \attackname requires that the adversary must be able to front-run a victim transaction, mitigating front-running can prevent an \attackname attack. The literature explores a variety of front-running mitigation solutions for the time-based transaction order-fairness~\cite{kelkar2020order,kursawe2021wendy,kelkar2021order,kelkar2021themis,zhang2020byzantine,cachin2021quick}. Although such fair ordering solutions could in principle minimize the threat of \attackname, they require fundamental changes in either the consensus or the application layer. In particular,~\cite{kelkar2020order} necessitates modifications to the underlying blockchain, whereas~\cite{kelkar2021themis,zhang2020byzantine} introduce an additional permissioned committee responsible for transaction ordering and require a \DeFi application redesign.


\begin{table*}[t]
    \centering
    \caption{Systematization of related work on automated smart contract exploitation.}
    \resizebox{\textwidth}{!}{%
    \begin{tabular}{@{}cccccccccc@{}}
    \toprule
    \multirow{3}{*}{\makecell{Attack/\\Exploit Generation Tool}} & \multirow{3}{*}{\makecell{Assumed Prior Knowledge}} & \multicolumn{2}{c}{Contract Vulnerability} & & \multicolumn{2}{c}{Exploit Generation} & & \multicolumn{2}{c}{Practicality} \\\cmidrule{3-4}\cmidrule{6-7}\cmidrule{9-10}
         & & \makecell{Searchspace Unrestricted\\From Vulnerability Patterns} &  \makecell{Cross-Contract\\Vulnerabilities} & & \makecell{Synthesize\\Transactions} & \makecell{Synthesize\\Contracts}  & & \makecell{Real-Time\\Capable (avg)} & \makecell{Application\\Agnostic} \\\midrule

      \multicolumn{10}{l}{Reward based exploit generation (the goal is to extract financial revenue, implicitly capturing vulnerabilities)} \\ \midrule
      \textbf{\attackname}  & Exploit transaction &  \hc~(limited by \txv)\dag & \fc  && \fc & \fc && \fc\ (\timeConsumptionMean s) & \fc \\
      Naive Imitation~\cite{qin2022quantifying}   & Exploit transaction & \hc~(limited by \txv)\dag & \fc && \fc & \ec && \fc\ (\naiveTimeConsumptionMean s) & \fc\\
      \textsc{DeFiPoser}~\cite{zhou2021just} & \DApp Model & \hc~(limited by \DApp models) & \fc && \fc & \ec && \hc\ ($5.93$ s) & \ec \\ \midrule
    
      \multicolumn{10}{l}{Software pattern exploit generation (the goal is to match/find predefined known vulnerability patterns)} \\ \midrule
      \textsc{ContractFuzzer}~\cite{jiang2018contractfuzzer}   & Patterns \& Contract \ABI & \ec & \ec && \fc & \ec & & \ec & \fc \\ 
      \textsc{Mythril}~\cite{consensy2017mythril}   & Patterns & \ec & \ec && \fc & \ec && \ec & \fc \\ 
      \textsc{teEther}~\cite{krupp2018teether}   & Patterns & \ec & \ec && \fc & \ec & & \ec & \fc \\ 
      \textsc{Maian}~\cite{nikolic2018finding} & Patterns & \ec & \ec && \fc & \ec && \fc\ ($10$ s) & \fc \\ 
      \textsc{Solar}~\cite{feng2020summary}   & Patterns \& Contract ABI\ & \ec & \ec && \ec & \fc & & \fc\ ($8$ s) & \fc \\ 
      \bottomrule
    \end{tabular}%
    }
    \begin{tabular}{@{}p{\textwidth}@{}}
    \footnotesize
    \dag While not designed to discover new vulnerability patterns, Section~\ref{sec:analysis} shows how light manual inspection of successful \txc  identifies new patterns.
    \end{tabular}
    \label{tab:taxonomy}
\end{table*}

\paragraph{Code Obfuscation}
Mature blockchains and \DeFi applications are not trivial to patch and redesign. There is a need for lightweight workarounds until front-running is mitigated in a more systematic manner. Consequently, one promising solution to the problem is to develop a bytecode-level obfuscation technique. Although code obfuscation has been utilized as a defense mechanism for decades~\cite{collberg1997obfuscation,hosseinzadeh2018obfuscation}, it has not yet been investigated as a countermeasure for smart contract attacks and vulnerabilities. Control flow obfuscation approaches~\cite{pawlowski2016probfuscation}, which aim to obscure the flow of a program to make it difficult for dynamic analysis to reason about the code, could be applied. Recall that, to copy the execution path, \attackname modifies the tainted basic blocks and hard-codes the code jumps (cf.\ Section~\ref{sec:smart-contract-synthesis}). If both the \emph{true} and \emph{false} branch of a tainted basic block are visited in a victim transaction, synthesizing an adversarial contract becomes difficult, because a hard-coded jump can only visit one branch. An \attackname-specific obfuscation scheme then could be designed so that hard-coded code jumps would result in an infinite loop.
Despite the fact that the obfuscation techniques are easy to adopt and provide basic protection against \attackname, they still have limitations. The primary issue is that sophisticated deobfuscation techniques could be developed to bypass those defenses. An obfuscation defense would also result in more complex contracts, increasing the execution and deployment gas costs. We leave the complete design and evaluation for future work.





\section{Related Work}
\paragraph{Exploit Tool Systematization}
We systematize the closest related work in Table~\ref{tab:taxonomy}, focussing on automated exploit generation tools. We distinguish among \emph{(i)} tools that aim to extract financial value and capture vulnerabilities in that process~\cite{zhou2021just, qin2022quantifying}, \emph{(ii)} tools detecting pre-defined vulnerability patterns within smart contracts~\cite{jiang2018contractfuzzer, consensy2017mythril, krupp2018teether, nikolic2018finding, feng2020summary}.

For reward-based exploitation tools, we find that the naive imitation method~\cite{qin2022quantifying} outperforms \attackname in terms of real-time performance due to the simplicity of the string replacing approach.
However, over the same evaluation timeframe (cf.\ Section~\ref{sec:evaluation}), \attackname captures higher financial gains and more \DeFi attacks compared to the naive method. \textsc{DeFiPoser}~\cite{zhou2021just} leverages an SMT solver to identify profitable strategies that can potentially unveil new vulnerability patterns. Nonetheless, the effectiveness of \textsc{DeFiPoser} relies on the precise modeling of blockchain applications, which requires substantial manual efforts. It is worth noting that the real-time performance of \textsc{DeFiPoser} hinges on the underlying SMT solver as well as the complexity of the mathematical models.

Tools that detect pre-defined vulnerability patterns within smart contracts take these patterns as input to conduct security analyses. Therefore, such tools may not discover different, or new vulnerability patterns. Yet, \textsc{Maian} and \textsc{Solar}, are designed to be real-time capable. \textsc{Solar} is the only tool prior to \attackname we are aware of which synthesizes contracts. \textsc{Solar}, however, requires access to a smart contract's \ABI. Recall that \ABI's may not be accessible for closed-source smart contracts. To our understanding, \textsc{Solar} focuses on vulnerabilities in a single contract and is therefore unlikely to capture composable (i.e., cross-contract) \DeFi attacks.

\paragraph{Smart Contract Attacks and Security}
Atzei et al.~\cite{atzei2017survey} provide a survey on attacks against smart contracts. Various papers focus on automatically finding exploits for vulnerable smart contracts by detecting vulnerabilities and generating malicious inputs~\cite{krupp2018teether,feng2019smartcopy,nikolic2018finding}. 
Qin et al.~\cite{qin2021attacking} explore \DeFi attacks through flash loans. Wu et al.~\cite{wu2021defiranger} propose a platform-independent way to recover high-level \DeFi semantics to detect price manipulation attacks.

\textit{Static analysis} tools~\cite{brent2018vandal,kalra2018zeus,grech2018madmax,tikhomirov2018smartcheck} such as Securify~\cite{tsankov2018securify}, Slither~\cite{feist2019slither}, and Ethainter~\cite{brent2020ethainter} detect specific vulnerabilities in smart contracts. Typically such tools achieve completeness but also report false positives.
Another common approach to detect smart contract vulnerabilities is to employ \textit{symbolic execution}~\cite{frank2020ethbmc,mossberg2019manticore,permenev2020verx}. Mythril~\cite{consensy2017mythril} and Oyente~\cite{luu2016oyente} use SMT-based symbolic execution to check EVM bytecode and simulate a virtual machine for execution-path exploration. Other tools~\cite{jiang2018contractfuzzer,grieco2020echidna,nguyen2020sfuzz,wustholz2020harvey,wustholz2020targeted} employ \textit{fuzzing techniques} to detect various vulnerabilities. Instead of employing fuzzing, \attackname instruments the \EVM to dynamically analyze smart contracts in one pass. Recently, bytecode rewriting for \textit{patching} smart contracts has been explored with \textsc{smartshield}~\cite{zhang2020smartshield} and sGuard~\cite{nguyen2021sguard}. \textsc{evmpatch}~\cite{rodler2021evmpatch} can integrate many static analysis tools to detect vulnerabilities and automate the whole lifecycle of deploying and managing an upgradeable contract.

\section{Conclusion}
The generalized blockchain imitation game is a new class of attacks on smart contract blockchains. Such imitations could be adopted for either selfish outcomes, e.g., a miner adopting \attackname can appropriate millions of USD worth of \DeFi attacks; or, on a brighter note, miners could help defend the \DeFi ecosystem as whitehat hackers, by front-running attacks and possibly refunding the resulting revenue to their victim. In this work, we show that such imitation games are practical and can yield significant value on Ethereum and \BSC.

\section*{Acknowledgments}
We thank the anonymous reviewers for the thorough reviews and helpful suggestions that significantly strengthened the paper. This work is partially supported by Lucerne University of Applied Sciences and Arts, and the Algorand Centres of Excellence programme managed by Algorand Foundation. Any opinions, findings, and conclusions or recommendations expressed in this material are those of the author(s) and do not necessarily reflect the views of the institutes.

{\footnotesize
\bibliographystyle{plain}
\bibliography{references.bib}
}
\appendix

\section{Naive Imitation Attack Example}\label{app:naive-replay-attack-example}
In this section, we show a real-world Ethereum transaction,\footnote{Transaction hash: \etherscanTx{0xe58214cfb38650089ce6bace5669a58e03557935ab8480467ae511df69ca40db}.} which is vulnerable to the naive imitation attack (cf.\ Listing~\ref{lst:replay_example}). This transaction invokes the function \texttt{increaseAllowance} of an ERC20 token contract GANGSINU\footnote{Address: \etherscanAddress{0x9796Bcece6b6032deB6f097b6F1cc180aE974feC}.} and mints one quadrillion of the GANGSINU token to the transaction sender. At the time of this transaction, one quadrillion of GANGSINU could be exchanged to~$\numprint{36.6}$ ETH on Uniswap.\footnote{Address: \etherscanAddress{0xAf852a23ee89999787146f8b4B440380E5Fac414}.} Note that the function \texttt{increaseAllowance} allows any address to mint an arbitrary amount of GANGSINU. When observing this transaction, a transaction imitation attacker could have called the same contract, front-run the transaction, and atomically exchanged the one quadrillion of GANGSINU to ETH on Uniswap, receiving a revenue of~$\numprint{36.6}$ ETH.

\begin{lstlisting}[float,floatplacement=H,label=lst:replay_example,language=Solidity, caption={Solidity code snippet of contract \href{https://etherscan.io/address/0x9796Bcece6b6032deB6f097b6F1cc180aE974feC\#code}{0x9796Bcece6b6032deB6f097b6F1cc180aE974feC}. The function \texttt{increaseAllowance} allows minting an arbitrary amount of the GANGSINU token to the \texttt{spender} address}.]
contract GANGSINU {
  function increaseAllowance(
    address spender,
    uint256 addedValue
  ) public virtual returns (bool) {
    _approve(
      _msgSender(),
      spender,
      _allowances[_msgSender()][spender] + addedValue
    );
    _mint(spender, addedValue);
    return true;
  }
}
\end{lstlisting}

\begin{table}[t]
    \centering
    \caption{Ethereum DeFi attacks and contract vulnerabilities identified by \attackname from the top-\AnalyzedVictims profitable victim transactions (ordered by USD profit).}
    \resizebox{\columnwidth}{!}{%
    \begin{tabular}{cclc}
    \toprule
    \makecell{Date\\(Block Number)} & \makecell{\attackname Profit\\(USD)} & Description & \makecell{Observed\\on \PtoP} \\
    \midrule
    Aug-03-2021 (\href{https://etherscan.io/tx/0xcd7dae143a4c0223349c16237ce4cd7696b1638d116a72755231ede872ab70fc}{12955063}) &	\PopsicleProfit & 	\faStar~Popsicle Finance & \tk \\
    Dec-15-2021 (\href{https://etherscan.io/tx/0x840e7bca49fd3999a3ec40a98915c726c9518d90469f5f3253eadb83d03168d8}{13810360}) &	$19.70$M &	\faBug~massDeposit & \tk \\
    Dec-11-2021 (\href{https://etherscan.io/tx/0x90a5d10ca9092a82464fd87efb1a7ed06c0d0420fde0cbb20af6e7ec9046c303}{13786402}) &	$19.12$M &  \faStarO~Sorbet Finance & \cx \\
    Apr-30-2022 (\href{https://etherscan.io/tx/0x2b023d65485c4bb68d781960c2196588d03b871dc9eb1c054f596b7ca6f7da56}{14684307}) &	$9.71$M &	\faStar~Saddle Finance & \tk \\
    Aug-10-2021 (\href{https://etherscan.io/tx/0x597d11c05563611cb4ad4ed4c57ca53bbe3b7d3fefc37d1ef0724ad58904742b}{12995895}) &	$5.69$M &	\faStarO~Punk Protocol & \tk \\
    Apr-30-2022 (\href{https://etherscan.io/tx/0x9549c0cb48ec5a5a2c4703cbbbbea5638028b2d8c8adc103220ef1c7fe5e99a3}{14684434}) &	$4.01$M &	\faStarO~Saddle Finance & \cx \\
    Oct-14-2021 (\href{https://etherscan.io/tx/0xbde4521c5ac08d0033019993b0e7e1d29b1457e80e7743d318a3c27649ca4417}{13417956}) &	$3.58$M &	\faStar~Indexed Finance & \tk \\
    Nov-27-2021 (\href{https://etherscan.io/tx/0x8233d0e7185a3bc7f4c3b19fa00d0ffe751f07b723dc3f6aa7e06c68401b6dd7}{13695970}) &	$2.04$M &	\faStarO~dYdX deposit vulnerability & \cx \\
    Dec-17-2021 (\href{https://etherscan.io/tx/0xa89d226e2b9e57a1bec3609b562693e5ebb23cd636a11dca8b43d2359feef341}{13822426}) &	$2.02$M &	\faBug~massDeposit & \tk \\
    Dec-15-2021 (\href{https://etherscan.io/tx/0xce4d7553fe87c8d029e37c3143f49bb550b60c93ec5591be95d2b6011713248b}{13810426}) &	$1.94$M &	\faBug~massDeposit & \tk \\
    Dec-15-2021 (\href{https://etherscan.io/tx/0xc939074c9b4d5f67ae69b66d29b591255603a53af54b65d4f1c3fac60b565c10}{13810215}) &	$1.80$M &	\faBug~massDeposit & \tk \\
    Apr-30-2022 (\href{https://etherscan.io/tx/0xe7e0474793aad11875c131ebd7582c8b73499dd3c5a473b59e6762d4e373d7b8}{14684518}) &	$1.58$M &	\faStar~Saddle Finance & \tk \\
    Dec-15-2021 (\href{https://etherscan.io/tx/0xc2317677605feac7133db5c2ffe9cbce18c09c882b66e0c9c91c0083f6b30d6c}{13810374}) &	$1.51$M &	\faBug~massDeposit & \tk \\
    Jun-16-2022 (\href{https://etherscan.io/tx/0x958236266991bc3fe3b77feaacea120f172c0708ad01c7a715b255f218f9313c}{14972419}) &	$1.08$M &   \faStar~Inverse Finance  & \tk \\
    Sep-15-2021 (\href{https://etherscan.io/tx/0xf3158a7ea59586c5570f5532c22e2582ee9adba2408eabe61622595197c50713}{13229001}) &	$1.07$M &	\faStar~NowSwap & \tk \\
    Dec-15-2021 (\href{https://etherscan.io/tx/0x900bebffb5bae5c1fb857dc2adeb7556d1da6d86675c3120e26c379e706ccb14}{13810401}) &	$990.50$K &	\faBug~massDeposit & \tk \\
    Jan-19-2022 (\href{https://etherscan.io/tx/0xe50ed602bd916fc304d53c4fed236698b71691a95774ff0aeeb74b699c6227f7}{14037237}) &	$862.22$K &	\faStarO~Multichain vulnerability & \cx \\
    Mar-26-2022 (\href{https://etherscan.io/tx/0x2e7d7e7a6eb157b98974c8687fbd848d0158d37edc1302ea08ee5ddb376befea}{14460636}) &	$717.80$K &	\faUserTimes~Auctus & \tk \\
    Jul-28-2022 (\href{https://etherscan.io/tx/0xa2704042b30a13b4d3c1b32fe11a523b78b47f0e8a5826931f397eaf2aa73ede}{15233205}) &	$695.63$K &	\faBug~Unautheticated Asset Redemption& \cx \\
    Mar-27-2022 (\href{https://etherscan.io/tx/0x613b2de3bb9043884a219296eeb1ada8c47b5a0262b9c68ca06ffd2de3a5d9f5}{14465382}) &	$657.98$K &	\faStar~Revest Finance & \cx \\
    Jan-22-2022 (\href{https://etherscan.io/tx/0xd59f1c77cad53d32024cde1a4e430f8ad268974febc107ec03f049ba0a4cf048}{14052155}) &	$519.69$K &	\faStar~Multichain vulnerability & \tk \\
    Nov-27-2021 (\href{https://etherscan.io/tx/0x33d0895cdf98bfc3d516b2e7a0de11014f254ec4f2861113d4afe85304bf1cd6}{13696312}) &	$495.51$K &	\faStarO~dYdX deposit vulnerability & \tk \\
    Aug-30-2021 (\href{https://etherscan.io/tx/0x55692ccc8ccf81b155044ed4109155ec7714dfae541fe4c4be23da8b18240248}{13124663}) &	$461.29$K &	\faStar~Cream Finance & \tk \\
    Dec-15-2021 (\href{https://etherscan.io/tx/0xf8248bb59b8d749d7fed9fd4ec624aed0bcd67d5ec6992bc00d5991892ebac86}{13810162}) &	$442.15$K &	\faBug~massDeposit & \tk \\
    Sep-14-2021 (\href{https://etherscan.io/tx/0xee9c92684464c5435243ad8fd4e048a9af2ce7338194ec68b4f610d72c0cae6c}{13222312}) &	$433.75$K &	\faBug~Faulty Authentication & \tk \\
    Aug-30-2021 (\href{https://etherscan.io/tx/0x77e2d72bd9d94f20b051f0a629a79e9a316b3ee3c6bf7495236f43bc24d379d8}{13124635}) &	$392.94$K &	\faStar~Cream Finance & \tk \\
    Aug-30-2021 (\href{https://etherscan.io/tx/0x8b4ec34be08527e549b7fc4863f23ca8a9c65824ac62a4327bd803f8cbb83fc2}{13124729}) &	$374.93$K &	\faStar~Cream Finance & \tk \\
    Feb-19-2022 (\href{https://etherscan.io/tx/0x3cd444604bfde45430ebbae516eac33a454fb6481e9f7ed2d9c4474d6c55bb49}{14234350}) &	$270.39$K & \faStarO~RigoBlock & \cx \\
    Aug-30-2021 (\href{https://etherscan.io/tx/0x51fd83401f2be2e16fed9c02c7f7df683e1f0c4bcc7cdccf58a1d810824c4992}{13124682}) &	$262.65$K &	\faStar~Cream Finance & \cx \\
    Aug-30-2021 (\href{https://etherscan.io/tx/0xf5a3225fd62ed183af9df48dd9b725727f8975d251165b40972cf54b3198fd70}{13124700}) &	$238.17$K &	\faStar~Cream Finance & \cx \\
    Oct-14-2021 (\href{https://etherscan.io/tx/0x6e5ad9ee58813c8e37aeeed62596124a189d1daff1f9101c565a20f4e24eb363}{13418167}) &	$215.28$K &	\faBug~Faulty Authentication & \tk \\
    Nov-26-2021 (\href{https://etherscan.io/tx/0xca68269685524d3818c98cb588c00a215fcc8a15c739c0a4468e078b3f3f3a7a}{13687922}) &	$208.02$K &	\faStar~Visor Finance & \tk \\
    Dec-15-2021 (\href{https://etherscan.io/tx/0xcd43b44826ed94ac969691d71e1eb60613c133905e609397f6d9433bc889eeb4}{13810232}) &	$173.32$K &	\faBug~massDeposit & \tk \\
    Jan-21-2022 (\href{https://etherscan.io/tx/0x640a215ac2dc32d33f0100a64424f3c2261e678a126d37a5e436209182740e49}{14051020}) & $134.90$K &	\faStar~Multichain vulnerability & \tk \\
    Aug-31-2021 (\href{https://etherscan.io/tx/0x4543c28c2ee41a8648d3cec96b25ca3e5be0ca78c7aec7f314601a486db5e4ee}{13130729}) &	$128.90$K &	\faBug~Faulty Authentication & \tk \\
    Aug-30-2021 (\href{https://etherscan.io/tx/0x0016745693d68d734faa408b94cdf2d6c95f511b50f47b03909dc599c1dd9ff6}{13124591}) &	$116.95$K &	\faStar~Cream Finance & \tk \\
    Jan-19-2022 (\href{https://etherscan.io/tx/0x2db9a6a51604e2be8b2c3469773afb201f0b48a318fb7e5f5e49175e818df5ba}{14036895}) &	$111.10$K &	\faStarO~Multichain vulnerability & \cx \\
    Dec-22-2021 (\href{https://etherscan.io/tx/0xea8759bc81333e0b7904d104e9cf2f83b0df772fafe733ac7647d042d72112eb}{13857734}) &	$109.62$K &	\faBug~Faulty Authentication & \cx \\
    Mar-27-2022 (\href{https://etherscan.io/tx/0x19b10c6d38f0b911fdc0e722d681a70a56699d70559eefef3d4d6fe88276c813}{14465427}) &	$108.96$K &	\faStar~Revest Finance & \cx \\
    Jan-21-2022 (\href{https://etherscan.io/tx/0x986ef252ddaed537df325efa31221dd0b0908da002dd040b799336a3a81df67c}{14046993}) &	$99.84$K &	\faStar~Multichain vulnerability & \cx \\
    Jan-18-2022 (\href{https://etherscan.io/tx/0x9600fea499fdbb708d1669b725eb693e08371a883d01054e66cfca13ff925f35}{14031416}) &	$80.08$K &	\faStarO~Multichain vulnerability & \cx \\
    Dec-12-2021 (\href{https://etherscan.io/tx/0x226247274854dc77f11602bd11a092674a47103d5455ce6e39785228aad21c5f}{13791112}) &	$79.28$K &	\faStarO~Sorbet Finance vulnerability & \cx \\
    Dec-12-2021 (\href{https://etherscan.io/tx/0xe0166e0c8f9815db39f7bbc7432afe0eb1b6489a1cfbdfdc42c226e6a8f31b9b}{13791101}) &	$79.26$K &	\faStarO~Sorbet Finance vulnerability & \cx \\
    Apr-13-2022 (\href{https://etherscan.io/tx/0xe12ae015c8023bbe6405662a3ddf5e8e106e7f6255e905b7312dcf65b27d755c}{14575370}) &	$78.05$K &	\faUserTimes~BasketDAO & \cx \\
    Jan-10-2022 (\href{https://etherscan.io/tx/0x2bf2a3ccfba747f042f38ad3ae40903fa5620db3913452dda5a149526b532d00}{13975768}) &	$72.84$K &	\faBug~Faulty Authentication & \cx \\
    \bottomrule
    \end{tabular}%
    }
    \resizebox{\columnwidth}{!}{%
    \begin{tabular}{@{}lll@{}}
    \\
    \faStar~---~Known DeFi Attacks & & \faUserTimes~---~Known Vulnerabilities \\
    \faStarO~---~Known whitehat DeFi Attacks & & \faBug~---~Newly Found Vulnerabilities
    \end{tabular}
    }
    \label{tab:attacks-and-vulnerabilities}
\end{table}

\begin{table}[t!]
    \centering
    \caption{\BSC DeFi attacks and contract vulnerabilities identified by \attackname from the top-\AnalyzedVictims profitable victim transactions (ordered by USD profit).}
    \resizebox{0.95\columnwidth}{!}{%
    \begin{tabular}{ccl}
    \toprule
    Date (Block Number) & \attackname Profit (USD) & Description \\
    \midrule
    Apr-12-2022 (\href{https://bscscan.com/tx/0xec317deb2f3efdc1dbf7ed5d3902cdf2c33ae512151646383a8cf8cbcd3d4577}{16886439}) & $11.52$M & 	\faStar~Elephant Money\\
    Aug-16-2021 (\href{https://bscscan.com/tx/0x7e2a6ec08464e8e0118368cb933dc64ed9ce36445ecf9c49cacb970ea78531d2}{10087724}) & $5.17$M & 	\faStar~XSURGE\\
    Dec-01-2021 (\href{https://bscscan.com/tx/0x85778af13373250cd7d2a09903128c086e76bbbb5adc61b3df74ae8b126abfd8}{13099703}) & $1.06$M & 	\faStar~CollectCoin\\
    Apr-09-2022 (\href{https://bscscan.com/tx/0xa5b0246f2f8d238bb56c0ddb500b04bbe0c30db650e06a41e00b6a0fff11a7e5}{16798807}) & $561.43$K & 	\faStar~Gymdefi\\
    Mar-20-2022 (\href{https://bscscan.com/tx/0x518411e06e276ad576ee52b8631b5bcc91bbf280aa3edf79e58017b3cecb0a8b}{16221156}) & $494.18$K & 	\faStar~TTS DAO\\
    Jan-17-2022 (\href{https://bscscan.com/tx/0xf8faed035c747331c623d40906d0142925829f6202013976c1ec8afc9eadb99a}{14433926}) & $460.99$K & 	\faStar~Crypto Burgers\\
    Jan-07-2022 (\href{https://bscscan.com/tx/0xa3d5b29a4447319e43424c9e1548d42a4f29498a7cda8af0461ab7a32a087a27}{14161877}) & $368.61$K & 	\faBug~massDeposit\\
    Nov-23-2021 (\href{https://bscscan.com/tx/0x7fe46c2746855dd57e18f4d33522849ff192e4e26c74835799ba8dab89099457}{12886417}) & $340.51$K & 	\faStar~Ploutoz Finance\\
    Dec-14-2021 (\href{https://bscscan.com/tx/0x9f4247837100201d8af77238f5b08c9bc572c5d42427e6b66898641e86c0e236}{13478895}) & $326.92$K & 	\faBug~massDeposit\\
    Jan-17-2022 (\href{https://bscscan.com/tx/0x56aece43604bde95aa27abc07c225288541cf07ac2c5bf5085fc8b82f905d474}{14433715}) & $289.23$K & 	\faStar~Crypto Burgers\\
    Oct-01-2021 (\href{https://bscscan.com/tx/0x24180e59f48bb6291213c3960ad516c23701e9d501fd6105f4087789f0a8d74a}{11406815}) & $270.06$K & 	\faStar~Twindex\\
    Aug-16-2021 (\href{https://bscscan.com/tx/0x4d691249665f8bfe18b89a5b7cb9133cc85afd68e7f9b3b876d9542067bc3dc9}{10090827}) & $244.50$K & 	\faStar~XSURGE\\
    Oct-02-2021 (\href{https://bscscan.com/tx/0x153a0d0376579dab66e067f3c655506793e0373b4c39eaab6144f083dc7b1bd6}{11410860}) & $237.55$K & 	\faStar~Twindex\\
    Apr-29-2022 (\href{https://bscscan.com/tx/0x8e47bdfbf0ba217498beb985418d1e79c35d6404b031748933e0c26bcf3bf68a}{17361160}) & $219.37$K & 	\faStar~Legend LFW\\
    Nov-20-2021 (\href{https://bscscan.com/tx/0xe887550858dc34fe36432103883a8396550e5d24c443d518fa0a7486d7fb9e0c}{12795006}) & $147.39$K & 	\faBug~Faulty Authentication\\
    Jun-01-2022 (\href{https://bscscan.com/tx/0x927723660249253399e54c192a5f989ceacf46fbb967ab364d4405155539bec8}{18305386}) & $142.22$K & 	\faStar~CoFiXProtocol\\
    Oct-04-2021 (\href{https://bscscan.com/tx/0x61a7b8e5dc62f7dfdc61a95e924d28aad50776033b34e4c26e98600b2f630faf}{11469791}) & $127.29$K & 	\faStar~WEDEX\\
    Aug-16-2021 (\href{https://bscscan.com/tx/0x8c93d6e5d6b3ec7478b4195123a696dbc82a3441be090e048fe4b33a242ef09d}{10090725}) & $125.40$K & 	\faStar~XSURGE\\
    Aug-12-2021 (\href{https://bscscan.com/tx/0x550b4403f240ea60ccc0ab2b7b92b362ad7f15a6dcb66748c92f9549dcee0095}{9969958}) & $99.72$K & 	\faStar~Maze Protocol\\
    Aug-12-2021 (\href{https://bscscan.com/tx/0xb0a0a995e1c29d8889f69561e6c4af1a71bd85514f32ae693725684f840c6711}{9969953}) & $99.68$K & 	\faStar~Maze Protocol\\
    Aug-12-2021 (\href{https://bscscan.com/tx/0x82083aa9f0bbd80c2d0d2116b71ba76791977133ef5a6a1ecdc69ce9e21848c9}{9970577}) & $99.62$K & 	\faStar~Maze Protocol\\
    Feb-07-2022 (\href{https://bscscan.com/tx/0xe7c726ff4cfbf18e05d9189ffc180eb33107d898469b25613ef457a0e139da34}{15053929}) & $80.49$K & 	\faStar~EarnHub\\
    Nov-20-2021 (\href{https://bscscan.com/tx/0xc860af83fd1d82b9fcfcfcee1621613ffa3cb90f50b616f0126d810d6fb9b70e}{12810507}) & $70.50$K & 	\faStar~Formation.fi\\
    Feb-07-2022 (\href{https://bscscan.com/tx/0xb041b2ba13f7c8d155b2986a1c2329251ac98d6bde194546031ad929eab05e06}{15053947}) & $68.86$K & 	\faStar~EarnHub\\
    May-07-2022 (\href{https://bscscan.com/tx/0x3a44e0e1b6ec970d0d1c82b8d9d7da2f7a5f556c8df8b80dcf1da9e58ced9801}{17602189}) & $68.28$K & 	\faBug~Unautheticated Minting\\
    Jan-16-2022 (\href{https://bscscan.com/tx/0x750a17071f8d39cc8f75ba0492af7508d4a8498b55d294ac4df2ca583611c000}{14429051}) & $64.54$K & 	\faStar~Brokoli Network\\
    Nov-26-2021 (\href{https://bscscan.com/tx/0xf29a43eee8ecb84badac7c4a4a6c558504e7d946797d218cc41fff582b6b45cd}{12961048}) & $64.39$K & 	\faBug~Faulty Authentication\\
    Feb-07-2022 (\href{https://bscscan.com/tx/0xb041b2ba13f7c8d155b2986a1c2329251ac98d6bde194546031ad929eab05e06}{15053947}) & $51.58$K & 	\faStar~EarnHub\\
    Aug-12-2021 (\href{https://bscscan.com/tx/0x6970c70f473486b83d2ac0e034337236a28a2c1c76c5c8453fee68b65bc84b8d}{9969776}) & $49.90$K & 	\faStar~Maze Protocol\\
    Aug-12-2021 (\href{https://bscscan.com/tx/0x900da25f42802bde9e4db1c39a6d51d80f21fb366c31774fd9b3a53eb41f0a03}{9969782}) & $49.88$K & 	\faStar~Maze Protocol\\
    Dec-05-2021 (\href{https://bscscan.com/tx/0x8bdeac2148d11e0091168ac9dd5480f01047a9a4df205fe701902895934d08b6}{13216927}) & $49.04$K & 	\faBug~Unautheticated Asset Redemption\\
    Dec-05-2021 (\href{https://bscscan.com/tx/0xd9e3972c3db69fa2947707641785671182aaa78ac15ae37eb1ac9522cea4e03c}{13216684}) & $46.73$K & 	\faBug~Unautheticated Asset Redemption\\
    Mar-13-2022 (\href{https://bscscan.com/tx/0xbfec34cbb7ec8b6ad337c3213e8efd0a1aade4bfb6ae769f77b5e8b338ce8163}{16015031}) & $45.19$K & 	\faStar~Paraluni\\
    Dec-05-2021 (\href{https://bscscan.com/tx/0xa32d92705d086bc56f9de7b4c1914a1fae9a5a3b3b1c9e6f41a44fae15d4ea81}{13216787}) & $44.50$K & 	\faBug~Unautheticated Asset Redemption\\
    Mar-20-2022 (\href{https://bscscan.com/tx/0x8256763e6dfd688ed27e337613aad0ce0c6dfe1e8543cf98406b8eea1b395130}{16221176}) & $40.91$K & 	\faStar~TTS DAO\\
    Feb-18-2022 (\href{https://bscscan.com/tx/0x7b04f8add8752504388f81a2fa59356e16f72228074933e1c8e8ea100536ac17}{15369735}) & $37.90$K & 	\faBug~Unverified Stake\\
    Jan-20-2022 (\href{https://bscscan.com/tx/0xfb166ed65b53b408662655bd876e5c23eb9347ab12ecfe68b3ed82114caed906}{14534738}) & $35.04$K & 	\faStar~AstroBirdz\\
    Feb-07-2022 (\href{https://bscscan.com/tx/0x40e69064c70d7db8b2dcbad441da9a06a507f8f90959da3c2583242f89e01d3c}{15053951}) & $35.00$K & 	\faStar~EarnHub\\
    Nov-13-2021 (\href{https://bscscan.com/tx/0x8e5fa0f5408e305f118c4b1af9d39614362674964bdba8d76ed85e31eb321955}{12621548}) & $33.21$K & 	\faStar~Welnance\\
    Oct-06-2021 (\href{https://bscscan.com/tx/0x7c5de9f5bc8b3f725b9511eba0e18a7f8410b9b6379f142be6660770714a6475}{11530941}) & $31.32$K & 	\faStar~Welnance\\
    Dec-14-2021 (\href{https://bscscan.com/tx/0xbf2dfd03637a10cd8dafb7ca58f968397bcaf7c7ffa21b7887fdde32c5fa4f96}{13479338}) & $29.15$K & 	\faBug~massDeposit\\
    May-07-2022 (\href{https://bscscan.com/tx/0x8c96b3314e30cf62bdfd4f94df38a2f040e171e849208b328dcd4ac2cdbcb748}{17607316}) & $28.97$K & 	\faStar~bistroo\\
    Aug-16-2021 (\href{https://bscscan.com/tx/0x797b3c58d68e6961d34c6d2d1e34933679a1ccd4d81541869495a9d57c0ab357}{10090919}) & $28.76$K & 	\faStar~XSURGE\\
    Dec-05-2021 (\href{https://bscscan.com/tx/0x60cb3a91d657f4d13c55fd8a4a4ac66e9526d2b728429da2b5728c53336768dc}{13216448}) & $26.80$K & 	\faBug~Unautheticated Asset Redemption\\
    Jan-18-2022 (\href{https://bscscan.com/tx/0xa22f03a42a9265813e2490ba1595634e25c001e4ac66f2e91cba96e627c9e077}{14478911}) & $26.04$K & 	\faStar~Crosswise\\
    Sep-20-2021 (\href{https://bscscan.com/tx/0x380356c98cf71e9d8f9301d139af29357c3b5ceeec8a6985ce17208ca8234797}{11086847}) & $25.06$K & 	\faBug~Faulty Authentication\\
    Dec-14-2021 (\href{https://bscscan.com/tx/0x5fb267c3ce9d86d3246b1894dd1c975c165949e5e3503a7bec21f291adff1eda}{13478986}) & $22.81$K & 	\faBug~massDeposit\\
    Feb-07-2022 (\href{https://bscscan.com/tx/0x211666221cb39cb66bd09682d7a96200627a61424b340a672957018fb81820ec}{15053956}) & $19.98$K & 	\faStar~EarnHub\\
    Nov-24-2021 (\href{https://bscscan.com/tx/0x42a7bfe1484705e9016fe6ae6c0cd6d711d0392e910506e6f419086add771367}{12909045}) & $19.83$K & 	\faStar~Ploutoz Finance\\
    Jul-13-2022 (\href{https://bscscan.com/tx/0x7f183df11f1a0225b5eb5bb2296b5dc51c0f3570e8cc15f0754de8e6f8b4cca4}{19523981}) & $18.95$K & 	\faStar~SpaceGodzilla\\
    Feb-18-2022 (\href{https://bscscan.com/tx/0x432180c4c02ecdcce834570416cff730d6a11fa61e43891f253e4a490cbe6bb9}{15369807}) & $16.14$K & 	\faBug~Unverified Stake\\
    Feb-07-2022 (\href{https://bscscan.com/tx/0x326e78ace30f4bb554871b39a5f5040aa7bc700aa7b62da90698184095582555}{15053961}) & $14.29$K & 	\faStar~EarnHub\\
    Nov-13-2021 (\href{https://bscscan.com/tx/0xf7a9c59953763a57f412b2e45455e70192b44356c602f7c79ddbfa9cb05f440b}{12622514}) & $13.03$K & 	\faStar~Welnance\\
    Mar-26-2022 (\href{https://bscscan.com/tx/0x819d3e5f597fa85d53a4f11d598071aca5af3de10f9da25f5ac672801ea7bc98}{16405137}) & $12.21$K & 	\faBug~Faulty Authentication\\
    Dec-14-2021 (\href{https://bscscan.com/tx/0x28972a0481f062dc64de8ec61dc4ca8a485192dbd9b80918bd02e5b86748c4b1}{13478639}) & $12.06$K & 	\faBug~massDeposit\\
    Jun-20-2022 (\href{https://bscscan.com/tx/0x9f5b02cb1ce2d75ba457a2d152d89b6d3932ff057c03739a0071fb816e0ebab3}{18854194}) & $11.85$K & 	\faStar~Whale Finance\\
    \bottomrule
    \end{tabular}%
    }
    \resizebox{0.9\columnwidth}{!}{%
    \begin{tabular}{@{}lll@{}}
    \\
    \faStar~---~Known DeFi Attacks & & \faBug~---~Newly Found Vulnerabilities
    \end{tabular}
    }
    \label{tab:bsc-attacks-and-vulnerabilities}
\end{table}

\section{Implementation Details}\label{app:implementation-details}
\paragraph{Code Details}
While step~\Circled{2} and~\Circled{6} are reasonably straightforward, the remaining steps make up for the bulk of the engineering effort. More specifically, step~\Circled{1} requires building and storing relevant dynamic information while concretely executing a transaction. 
The goal of step~\Circled{3} is to discover all tainted blocks within one execution, as in to enable \attackname's real-time property. This constraint implies that we do not explore fuzzing or other iterative techniques. 
The challenge we have to overcome in step~\Circled{3}, is that when executing \txc, a tainted basic block often leads to a different execution path from \txv, resulting in an early termination, or failure (e.g., \texttt{STOP} and \texttt{REVERT}). This implies that given multiple tainted basic blocks executed in sequence, a tainted basic block $\mathsf{bb}_i$ may hinder the detection of a subsequent tainted basic block $\mathsf{bb}_{i+j}$.
We hence instrument the \EVM to allow manipulating the stack and forcing \txc to follow \txv's execution path. 
In step~\Circled{5}, we amend the bytecode of synthesized contracts to recover storage variables, redirect contract invocations, and fix jump destinations (cf.\ Section~\ref{sec:smart-contract-synthesis}).
\paragraph{Early abort}
Recall that in step~\Circled{4}, \adversary identifies all smart contracts that need to be replaced, i.e., \rscvi (cf.\ Section~\ref{sec:patch-identifier}). We notice that the replacement of an asset contract (e.g., USDC) creates a new asset with the same contract code but a different financial value. \adversary can abort \attackname, if at least one of \rscvi is such an asset contract. Note that this ``early abort'' is optional because \adversary can attempt to replace every \rscvi and validate the profitability in step~\Circled{6}. To increase our evaluation efficiency, we choose to abort \attackname once step~\Circled{4} detects an asset contract.

Multiple methods may allow to determine whether a contract is an asset contract. Instead of maintaining an asset contract dictionary, we apply a more practical method in our evaluation. We identify unique EVM logs (i.e., events) while executing \txv. We observe that most asset contract implementations follow a particular asset standard (e.g., ERC20). Whenever an asset contract receives an incoming invocation, the contract generates an event, identifiable by a \href{https://www.4byte.directory/}{unique hash value}. For example, to identify ERC20 contracts, \adversary checks if an event matches an ERC20 event. If there is a match, the contract generating such event is classified as an ERC20 contract. Our method therefore generalizes to any other asset standards, avoiding to regularly maintain a list of potential asset contracts. Note, however, that this method assumes that an adversary does not emit dummy events because the adversary could otherwise evade \attackname's attack. 

\section{Evaluation Analysis}\label{app:evaluation-analysis}
Table~\ref{tab:attacks-and-vulnerabilities} and~\ref{tab:bsc-attacks-and-vulnerabilities} list the DeFi attacks and contract vulnerabilities identified by \attackname from the top-\AnalyzedVictims profitable victim transactions on Ethereum and \BSC respectively.

We proceed to outline the newly found vulnerabilities.

\paragraph{Unverified Stake} The unverified stake vulnerability exists with an asset staking contract (\bscscanAddress{0x1E97D2363c6261D0ca4B182c7C670499afB93c73}) on \BSC. The contract allows a user to \emph{stake} asset without checking the financial value of the staked asset. An attacker hence can stake self-created assets (essentially a self-deployed smart contract) and immediately \emph{unstake} with other valuable assets from the staking contract, yielding a profit.

\paragraph{Unauthenticated Minting} A token contract (\bscscanAddress{0x0fa73D350E5e5bf63863f49Bb4bA3e87A20c93Fb}) on \BSC allows any account to mint an arbitrary amount of tokens and sell the minted tokens to profit.

\paragraph{Unauthenticated Asset Redemption} Contracts outlined in Table~\ref{tab:unauthenticated_asset_redemption} allow any account to redeem assets controlled by the contracts.

\begin{table}[h]
    \centering
    \caption{\attackname identifies two smart contracts with the unauthenticated asset redemption vulnerability.}
    \resizebox{\columnwidth}{!}{%
    \begin{tabular}{cl}
    \toprule
    Chain &  Contract Address \\\midrule
    Ethereum & \href{https://etherscan.io/address/0xbe5002a6b631570b0970838dcad7dc0aa2525282}{0xbE5002A6b631570b0970838dcad7dc0AA2525282} \\
    \BSC & \href{https://bscscan.com/address/0x521ef54063148e5f15f18b9631426175cee23de2}{0x521ef54063148E5F15F18B9631426175ceE23DE2} \\
    \bottomrule
    \end{tabular}%
    }
    \label{tab:unauthenticated_asset_redemption}
\end{table}

\paragraph{Faulty Authentication} We moreover identify eight vulnerable contracts (cf.\ Table~\ref{tab:faulty_authentication_contracts}) that are closed-source. By analyzing the corresponding transaction trace, we find that the vulnerabilities fall into the faulty authentication category (i.e., assets controlled by a vulnerable contract can be transferred without an authentication).
\begin{table}[h]
    \centering
    \caption{\attackname identifies eight closed-source smart contracts with the faulty authentication vulnerability.}
    \resizebox{\columnwidth}{!}{%
    \begin{tabular}{cl}
    \toprule
    Chain &  Contract Address \\\midrule
    \multirow{4}{*}{Ethereum} & \href{https://etherscan.io/address/0x37f4bf68bb295986d8a19e25528c441b5b4d4902}{0x37F4Bf68Bb295986d8a19E25528C441b5B4d4902} \\
    & \href{https://etherscan.io/address/0xf517a01ce955472d90cd0a8403629a46d7374e31}{0xf517A01CE955472d90cD0A8403629a46D7374E31} \\
    & \href{https://etherscan.io/address/0x35ba14ea6935ccbedcd745ed9096709d3e14d7b3}{0x35ba14eA6935cCbedCD745Ed9096709D3e14D7b3} \\
    & \href{https://etherscan.io/address/0x44cCdCD59984848a749e9f999B08F2b68153e123}{0x44cCdCD59984848a749e9f999B08F2b68153e123} \\\midrule
    \multirow{4}{*}{\BSC} & \href{https://bscscan.com/address/0xa6a2158f14f10b2288c914ee03710a5eb2bd3d7b}{0xA6a2158F14F10B2288c914eE03710a5eB2bd3d7b} \\
    & \href{https://bscscan.com/address/0x47e5a87c15a316040d34d1af075c8dc6bc0e63c7}{0x47E5a87C15a316040d34d1af075c8Dc6bC0e63c7} \\
    & \href{https://bscscan.com/address/0xafe1b53405667d098ec7b60c3794418e14a8fd21}{0xaFe1b53405667D098EC7b60C3794418E14A8fd21} \\
    & \href{https://bscscan.com/address/0x60093848a99e0589cdec170c941b6def5bac9b95}{0x60093848a99e0589cdec170C941B6Def5BAc9b95} \\
    \bottomrule
    \end{tabular}%
    }
    \label{tab:faulty_authentication_contracts}
\end{table}

\end{document}

%% file: languages.tex
\definecolor{codegreen}{rgb}{0,0.6,0}
\definecolor{codegray}{rgb}{0.5,0.5,0.5}
\definecolor{codepurple}{rgb}{0.58,0,0.82}
\definecolor{backcolour}{rgb}{0.95,0.95,0.92}
\usepackage{inconsolata}

\lstdefinelanguage{Solidity}{
	keywords=[1]{anonymous, assembly, assert, balance, break, call, callcode, case, catch, class, constant, continue, constructor, contract, debugger, default, delegatecall, delete, do, else, emit, event, experimental, export, external, false, finally, for, function, gas, if, implements, import, in, indexed, instanceof, interface, internal, is, length, library, log0, log1, log2, log3, log4, memory, modifier, new, payable, pragma, private, protected, public, pure, push, require, return, returns, revert, selfdestruct, send, solidity, storage, struct, suicide, super, switch, then, this, throw, transfer, true, try, typeof, using, value, view, while, with, addmod, ecrecover, keccak256, mulmod, ripemd160, sha256, sha3}, 
	keywordstyle=[1]\color{magenta}\bfseries,
	keywords=[2]{address, bool, byte, bytes, bytes1, bytes2, bytes3, bytes4, bytes5, bytes6, bytes7, bytes8, bytes9, bytes10, bytes11, bytes12, bytes13, bytes14, bytes15, bytes16, bytes17, bytes18, bytes19, bytes20, bytes21, bytes22, bytes23, bytes24, bytes25, bytes26, bytes27, bytes28, bytes29, bytes30, bytes31, bytes32, enum, int, int8, int16, int24, int32, int40, int48, int56, int64, int72, int80, int88, int96, int104, int112, int120, int128, int136, int144, int152, int160, int168, int176, int184, int192, int200, int208, int216, int224, int232, int240, int248, int256, mapping, string, uint, uint8, uint16, uint24, uint32, uint40, uint48, uint56, uint64, uint72, uint80, uint88, uint96, uint104, uint112, uint120, uint128, uint136, uint144, uint152, uint160, uint168, uint176, uint184, uint192, uint200, uint208, uint216, uint224, uint232, uint240, uint248, uint256, var, void, ether, finney, szabo, wei, days, hours, minutes, seconds, weeks, years},	
	keywordstyle=[2]\color{codepurple}\bfseries,
	keywords=[3]{block, blockhash, coinbase, difficulty, gaslimit, number, timestamp, msg, data, gas, sender, sig, value, now, tx, gasprice, origin},	
	keywordstyle=[3]\color{teal}\bfseries,
	identifierstyle=\color{black},
	sensitive=false,
	comment=[l]{//},
	morecomment=[s]{/*}{*/},
	commentstyle=\color{gray}\ttfamily,
	stringstyle=\color{red}\ttfamily,
	morestring=[b]',
	morestring=[b]"
}

\lstdefinelanguage{Decompiled}{
	keywords=[1]{anonymous, assembly, assert, balance, break, call, callcode, case, catch, class, constant, continue, constructor, contract, debugger, default, delegatecall, delete, do, else, emit, event, experimental, export, external, false, finally, for, function, gas, if, implements, import, in, indexed, instanceof, interface, internal, is, length, library, log0, log1, log2, log3, log4, memory, modifier, new, payable, pragma, private, protected, public, pure, push, require, return, returns, revert, selfdestruct, send, solidity, storage, struct, suicide, super, switch, then, this, throw, transfer, true, try, typeof, using, value, view, while, with, addmod, ecrecover, keccak256, mulmod, ripemd160, sha256, sha3}, 
	keywordstyle=[1]\color{magenta}\bfseries,
	keywords=[2]{address, bool, byte, bytes, bytes1, bytes2, bytes3, bytes4, bytes5, bytes6, bytes7, bytes8, bytes9, bytes10, bytes11, bytes12, bytes13, bytes14, bytes15, bytes16, bytes17, bytes18, bytes19, bytes20, bytes21, bytes22, bytes23, bytes24, bytes25, bytes26, bytes27, bytes28, bytes29, bytes30, bytes31, bytes32, enum, int, int8, int16, int24, int32, int40, int48, int56, int64, int72, int80, int88, int96, int104, int112, int120, int128, int136, int144, int152, int160, int168, int176, int184, int192, int200, int208, int216, int224, int232, int240, int248, int256, mapping, string, uint, uint8, uint16, uint24, uint32, uint40, uint48, uint56, uint64, uint72, uint80, uint88, uint96, uint104, uint112, uint120, uint128, uint136, uint144, uint152, uint160, uint168, uint176, uint184, uint192, uint200, uint208, uint216, uint224, uint232, uint240, uint248, uint256, var, void, ether, finney, szabo, wei, days, hours, minutes, seconds, weeks, years},	
	keywordstyle=[2]\color{teal}\bfseries,
	keywords=[3]{block, blockhash, coinbase, difficulty, gaslimit, number, timestamp, msg, data, gas, sender, sig, value, now, tx, gasprice, origin},	
	keywordstyle=[3]\color{codepurple}\bfseries,
	identifierstyle=\color{black},
	sensitive=false,
	comment=[l]{//},
	morecomment=[s]{/*}{*/},
	commentstyle=\color{codegreen}\ttfamily,
	stringstyle=\color{red}\ttfamily,
	morestring=[b]',
	morestring=[b]"
}

\lstdefinelanguage{Opcode}{
	keywords=[1]{REVERT,JUMPDEST,SELFBALANCE,CALLER,PUSH20,EQ,DUP1,PUSH1,PUSH2,PUSH4,JUMPI,POP,MSTORE,ADDRESS,GAS,STATICCALL,CALL},
	keywordstyle=[1]\color{teal}\bfseries,
	keywords=[2]{PC, Disassembled, Code},
	keywordstyle=[2]\color{codegray}\ttfamily,
	comment=[l]{//},
	morecomment=[s]{/*}{*/},
	commentstyle=\color{gray}\ttfamily,
}